\documentclass[preprint,aps]{revtex4-1}
\usepackage{graphicx}
\usepackage{hyperref}
\usepackage{epstopdf}
\usepackage{float}
\usepackage{bm}
\usepackage{amsmath}
\def\be{\begin{equation}}
\def\ee{\end{equation}}
\def\bea{\begin{eqnarray}}
\def\eea{\end{eqnarray}}
\def\ba{\begin{array}}
\def\ea{\end{array}}
\def\bc{\begin{center}}
\def\ec{\end{center}}
\def\la{\begin{langle}}
\def\ra{\end{rangle}}

\begin{document}
\title{Chiral Odd Generalized Parton Distributions and Spin Densities in the Impact Parameter Space}
\author{Narinder Kumar and Harleen Dahiya}
\affiliation{Department of Physics, Dr. B.R. Ambedkar National
Institute of Technology, Jalandhar, 144011, India}
\begin{abstract}
In the present work we have studied the chiral odd Generalized Parton Distributions (GPDs) in the impact parameter space by assuming a flexible parametrization in a quark-diquark model. In order to obtain the explicit contributions from the up and down quarks, we have considered  both the scalar (spin-0) and the axial-vector (spin-1) configurations for the diquark. We have also studied the spin densities for the up and down quarks in this model for monopole, dipole and quadrupole contributions for unpolarized and polarized quarks in unpolarized and polarized proton.
\end{abstract}
\maketitle

\section{Introduction}
Many experiments are presently running worldwide and some have finished taking
the data towards the study of hadronic structure. There has been
an enormous interest to understand the partonic distribution and many
models have been proposed to explain the hadronic properties theoretically. Last two
decades have been dedicated towards the study of generalized parton
distributions (GPDs) which contain 3-D structure information of the hadrons \cite{gpds,gpds1,gpds2,gpds3,gpds4,pire,miller1,miller2,miller3}.
Several experiments, for example, H1 collaboration \cite{h1coll,h1coll1}, ZEUS
collaboration \cite{zeus,zeus1} and fixed target experiments at HERMES
\cite{hermes} have completed taking data on deeply virtual Compton scattering
(DVCS). Experiments are also running at JLAB, Hall A and B \cite{clas} and
COMPASS at CERN \cite{compass}  to access the GPDs.

GPDs have been classified into two types: the chiral even GPDs where quark helicity does not flip ($H, E, \tilde{H}, \tilde{E}$) and the chiral odd GPDs which include the quark helicity flip ($H_T, E_T, \tilde{H}_T, \tilde{E}_T$). In a complete quark-parton model of the nucleon, the quark density or unpolarized
distribution $f_1(x)$ is the probability to find a quark with momentum fraction $x$ in the parent nucleon without considering the orientation of the spin,
$g_1(x)$ measures the helicity of quark in the longitudinally polarized nucleon and $h_1(x)$ is the number density of quarks having polarization
parallel to the nucleon minus the quarks anti parallel to nucleon polarization. These quark distributions can be respectively obtained from $H, \tilde{H}$ and $H_T$. GPDs depend
on three variables $x, \ \zeta, \ t$, where $x$ is the fraction of momentum
transferred, $\zeta$ (skewness) gives the longitudinal momentum transfer and
$t$ is the square of the momentum transfer in the process. However, it has to
be realized that only two of these variables $\zeta$ (fully defined by
detecting the scattered lepton $\zeta=x_b$, where $x_b$ is the Bjorken variable) and $t$ (fully defined by detecting either the recoil proton or
meson) are accessible experimentally.

Chiral even GPDs allow us to access partonic configurations not only with a
given longitudinal momentum fraction but also at a specific (transverse)
location inside the hadron. In the forward limit they reduce to usual parton densities and when integrated over $x$, they reduce to
the form factors which are the non-forward matrix elements of the current
operator and describe how the forward matrix element (charge) is distributed in
position space. They can be related to the angular
momentum carried by quarks inside the nucleon and the distribution of quarks can
be described in the longitudinal direction as well as in the impact parameter
space \cite{diehl1,burkardt,manohar,burkardt1,chakrabarty}.  On the other hand,
Fourier transform (FT) of the GPDs w.r.t. transverse momentum transfer
gives the distribution of partons in transverse position space \cite{dahiya}. Recently chiral even GPDs from DVCS amplitude for non-zero skewness in
longitudinal and impact parameter space have been studied \cite{kumar}. For the case of non-zero skewness,  both longitudinal and transverse distribution of partons is obtained in the hadron \cite{burkardt1} whereas for the case of zero skewness, the momentum transfer is only in the transverse direction thus giving the transverse distribution of the partons.  Chiral even GPDs encode the various properties of the hadrons for example electromagnetic form factors,
gravitational form factors \cite{brodsky,kumar1} and also provide the detailed information
upon the charge and magnetization densities \cite{miller,kumar2}.

The chiral odd GPDs, in the forward limit, reduce to transversity $h_1(x)$. When integrated over $x$, the fundamental term $2\tilde{H}_T+E_T$  has been studied   in a self-consistent two-body model \cite{burkardt2,burkardt3,dahiya_chiral_odd}, basically for the quantum fluctuation of an electron at one-loop in QED. This term is of great interest as it provides valuable information about the correlation between the spin and orbital angular momentum of the quarks inside the nucleon. There is however no direct interpretation for $\tilde{E}_T$ \cite{diehl}. Chiral odd GPDs have been studied in the longitudinal and transverse position spaces \cite{dipankar} where a field theory inspired model of spin- 1/2 system is considered. GPDs have also been discussed in a simple version of MIT bag model with an SU(6) proton wavefunction \cite{scopetta} and in light-front constituent quark models \cite{pincetti}.

In addition to this, relatively small number of studies have been done to study the nucleon spin densities which describe the quark distributions in
the nucleon for unpolarized and polarized quarks in unpolarized and polarized nucleon. The nucleon spin densities have been studied in light-front
constituent quark model where the first $x$-moments of spin densities have been
obtained for the up and down quarks \cite{pasquini}. Lattice calculations regarding the lowest
two $x$-moments of the transverse spin densities of the quarks in the nucleon
have also been performed \cite{gockeler} predicting that the Boer-Mulders function
$h_1^{\perp}$ is large and negative for both the up and down quarks. This is based
on the arguments given by the Burkardt \cite{burkardt2} where transverse
deformation of parton distribution has been discussed.
Recently, transverse distortion in impact parameter space has been studied  in the light-cone model \cite{kumar3}. Spin densities in transverse plane and generalized quark distributions have been studied in Ref. \cite{diehl} providing the relation between second leading twist
T-odd quark transverse momentum distributions, the Boer-Mulders distribution function $h_1^{\perp}$ and a
linear combination of GPDs.

In the present work, we have studied the transversely polarized chiral odd GPDs and the spin densities in the impact parameter space which are not so well known aspects of the nucleon structure. We have used the covariant model \cite{goldstein1} to evaluate the quark-proton helicity amplitudes. The formalism is based upon the dissociation of the initial proton into a quark and a fixed mass system (diquark). To obtain the distinct predictions for the up and down quarks we have considered both the spin-0 (scalar) and spin-1 (axial-vector) configurations for the diquark \cite{ahmad,ahmad1,jakob}. Further, we have obtained the results for the fundamental term $2 \tilde{H}_T + E_T$, linked with the transverse momentum distributions (TMDs) and it's first moment providing the proton's transverse anomalous magnetic moment. We have also studied the spin densities for monopole, dipole and quadrupole contributions for different situations, for example, when the quarks and proton both are unpolarized, when the quarks are polarized but proton is unpolarized and finally when both the quarks and proton are polarized but in different directions.

The plan of the paper is as follows. To make the manuscript
readable as well as to facilitate discussion, in Sec \ref{gpds} we present some of the essentials of the  chiral even and chiral odd GPDs. In Sec. \ref{impact}, the GPDs in the impact parameter space have been discussed. Section \ref{spin} includes the details of the spin densities. Section \ref{summ} comprises the summary and
conclusions.

\section{Chiral even and chiral odd Generalized Parton Distributions \label{gpds}}
The GPDs can be defined from the quark-quark proton correlator function as follows
\be
\Phi^{\Gamma}_{\Lambda', \Lambda}(x,\Delta, P)= \int \frac{dz^-}{2 \pi} e^{i x P^+ z^-} \langle p', \Lambda'| \overline{\psi}(-\frac{z}{2}) \Gamma \psi(-\frac{z}{2})| p, \Lambda \rangle|_{z^+=0, z_\perp =0},
\label{correaltor}
\ee
where $\Gamma= \gamma^+$, $\gamma^+ \gamma_5$, $i \sigma^{i +} \gamma_5$ $(i=1,2)$, with target spins $\Lambda$, $\Lambda'$ and momenta $p$, $p'$.\\
For the chiral odd case, we take $\Gamma= i \sigma^{i +} \gamma_5$. The correlator can be parametrized as
\bea
\Phi^{i \sigma^{i +} \gamma_5}_{\Lambda', \Lambda}(x,\zeta, t)&=& \overline{U}(P', \Lambda') \Big(i \sigma^{i +} H_T(x,\zeta,t)+ \frac{\gamma^+ \Delta^i - \Delta^+ \gamma^i}{2 M} E_T(x,\zeta,t)\nonumber\\
&& +\frac{P^+ \Delta^i-\Delta^+ P^i}{M^2} \tilde{H}_T(x,\zeta,t)+\frac{\gamma^+ P^i - P^+ \gamma^i}{2 M}\Big) U(P,\Lambda).
\label{correaltor}
\eea
The four momentum light-cone components in a asymmetric frame can be defined as:
\bea
P&=&\Big(P^+, \frac{M^2}{P^+},0\Big),\nonumber\\
P'&=& \Big((1-\zeta) P^+, \frac{M^2+\Delta_\perp^2}{(1-\zeta)P^+},\Delta_\perp \Big), \nonumber\\
\Delta &=& \Big(\zeta P^+,\frac{(1-\zeta/2)M^2+\Delta_\perp^2/2}{(1-\zeta)P^+},\Delta_\perp \Big), \nonumber\\
t&=&t_0 - \frac{\Delta_\perp^2}{(1-\zeta)}, t_0=- \frac{\zeta^2 M^2}{(1-\zeta)},
\eea
where $\Delta_\perp$ is the square momentum transfer in the process and $\zeta$ is the skewness parameter.

The helicity amplitude $f_{\Lambda_\gamma 0}^{\Lambda \Lambda'}$ for the deep virtual meson production (DVMP) can be introduced with the helicities of the virtual photon and the initial proton being $\Lambda_\gamma$, $\Lambda$ and the helicities of the pion and the proton being $0$, $\Lambda'$ respectively. Following Ref. \cite{goldstein0,goldstein}, the helicity amplitude $f_{\Lambda_\gamma 0}^{\Lambda \Lambda'}$ into $g^{\Lambda \Lambda'}_{\Lambda_\gamma 0}(x, \zeta, t , Q^2)$ and $A_{\Lambda' \lambda', \Lambda \lambda}(x, \zeta, t)$ can be decomposed into hard part and soft part as follows
\be
f^{\Lambda \Lambda'}_{\Lambda_\gamma 0}(\zeta, t)= \sum_{\lambda, \lambda'}g^{\Lambda \Lambda'}_{\Lambda_\gamma 0}(x, \zeta, t , Q^2) \otimes A^{S}_{\Lambda' \lambda', \Lambda \lambda}(x, \zeta, t),
\label{helicity_amplitudes}
\ee
where we have used the superscript S for denoting the spin of scalar and axial vector diquark contributions towards the GPDs. The convolution integral is given by $ \otimes \rightarrow \int_{-\zeta+1}^{1} dx$, the term $g^{\Lambda \Lambda'}_{\Lambda_\gamma 0}$ (hard part) describes the partonic subprocess  $\gamma^* + q \rightarrow \pi^0 + q$ and quark-proton helicity amplitude $A^S_{\Lambda' \lambda', \Lambda \lambda}$ (soft part) contains the GPDs. The model which we
used here is the quark-diquark model in which the proton dissociates into a
quark and a recoiling mass system which is considered as a diquark.

\begin{figure}
\includegraphics[width=12cm]{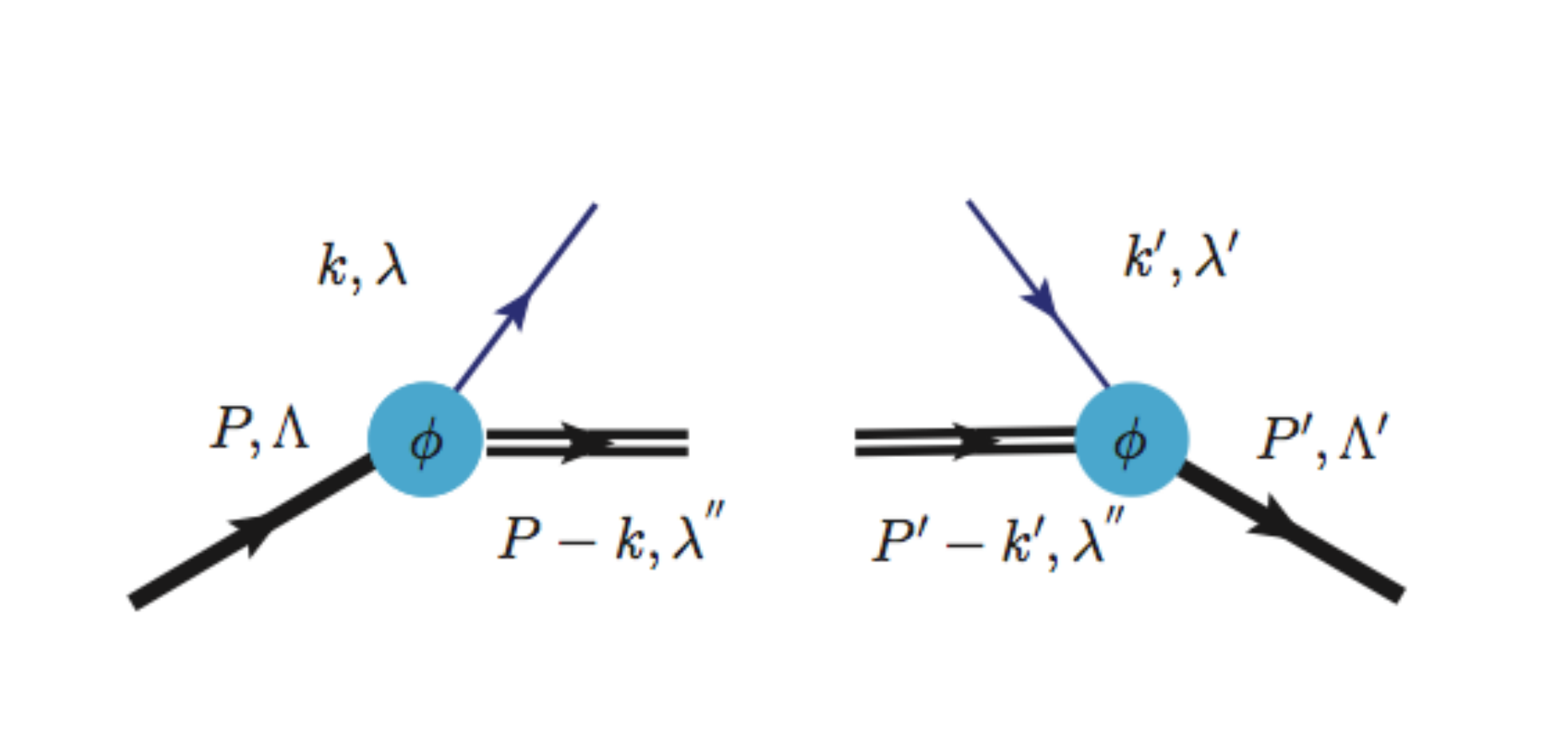}
\caption{Vertex tree level diagram defining the quark-diquark model.}
\label{modelfigure}
\end{figure}

The quark-proton scattering amplitudes at leading order with
proton-quark-diquark vertices can be computed from Fig. \ref{modelfigure}. We have
considered the spin-0 and spin-1 diquark so that we can obtain the explicit up and down quarks contributions by using the SU(4) symmetry of the proton wavefunction. For the case of spin-0 scalar diquark, the amplitude can be written as follows \cite{goldstein}
\be
A_{\Lambda' \lambda', \Lambda \lambda}^{0}= \int d^2 k_\perp \phi^*_{\Lambda' \lambda'}(k', P')\phi_{\Lambda \lambda}(k, P), \label{a0}
\ee
where the vertex functions can be defined as
\bea
\phi_{\Lambda, \lambda}(k, P)&=& \Gamma(k) \frac{\overline{u}(k,\lambda) U(P,\Lambda)}{k^2-m^2}, \nonumber\\
\phi^*_{\Lambda', \lambda'}(k', P')&=& \Gamma(k') \frac{\overline{U}(P', \Lambda')u(k', \lambda')}{k'^2-m^2}.
\eea
The $\Gamma(k)$ give the scalar coupling at proton-quark-diquark vertex and can be defined as
\be
\Gamma(k) = g_s \frac{k^2-m^2}{(k^2-M^2_\Lambda)^2},
\ee
where $g_s$ is a coupling constant.
The vertex structures for spin-0 diquark are given as
\bea
\phi^{*}_{++}(k,p)&=& \mathcal{A}(m+ M x),\nonumber\\
\phi^{*}_{+-}(k,p)&=& \mathcal{A} (k_1+ i k_2),\nonumber\\
\phi_{--}(k,p)&=&\phi_{++}(k,p),\nonumber\\
\phi_{-+}(k,p)&=& - \phi^{*}_{+-}(k,p).
\eea

For the case of spin-1 axial-vector diquark, the amplitude can be written as follows \cite{goldstein}
\be
A_{\Lambda' \lambda', \Lambda \lambda}^{1}= \int d^2 k_\perp \phi^{* \mu}_{\Lambda' \lambda'}(k', P')\sum_{\lambda''}\epsilon_\mu^{* \lambda''}\epsilon_\nu^{\lambda''}\phi^{\nu}_{\Lambda \lambda}(k, P), \label{a1}
\ee
where $\lambda''$ is the diquark helicity. In the present work we consider only the transverse helicities. Further, the vertex functions in this case can be defined as
\bea
\phi^\nu_{\Lambda, \lambda}(k, P)&=& \Gamma(k) \frac{\overline{u}(k,\lambda)\gamma^5 \gamma^\mu U(P,\Lambda)}{k^2-m^2}, \nonumber\\
\phi^*_{\Lambda', \lambda'}(k', P')&=& \Gamma(k') \frac{\overline{U}(P', \Lambda')\gamma^5 \gamma^\mu u(k', \lambda')}{k'^2-m^2}.
\eea
The explicit vertex structures for spin-1 diquark are given as
\bea
\phi^{+}_{++}(k,p)&=&\mathcal{A} \frac{k_1 - i k_2}{1-x}, \nonumber\\
\phi^{-}_{++}(k,p)&=&-\mathcal{A} \frac{(k_1 + i k_2) x}{1-x},\nonumber\\
\phi^{+}_{+-}(k,p)&=& 0, \nonumber\\
\phi^{-}_{+-}(k,p)&=& - \mathcal{A} (m+ M x), \nonumber\\
\phi^{+}_{-+}(k,p)&=&  - \mathcal{A} (m+ M x), \nonumber\\
\phi^{-}_{-+}(k,p)&=& 0,
\eea
where $\mathcal{A} = \frac{1}{\sqrt{x}} \frac{\Gamma{(k)}}{k^2-m^2}$, $k^2-m^2= M^2 x - \frac{x}{1-x} M_x^2-m^2- \frac{k_\perp^2}{1-x}$ and ${k'}_i=k_i - (1-x) \Delta_i$ $(i=1,2)$. Here $M_x$ is the invariant mass of the diquark and we have taken it at a
fixed value.

The chiral odd GPDs can now be expressed in terms of the helicity amplitudes and are given as
\bea
\tau[2 \tilde{H}_T(x,0,t)+ E_T(x,0,t)]^{S}&=& A^{S}_{++,+-}+A^{S}_{-+,--},\nonumber\\
H_T(x,0,t)^{S}&=& A^{S}_{++,--}+A_{-+,+-},\nonumber\\
\tau^2 \tilde{H}_T (x,0,t)^{S}&=& - A^{S}_{-+,+-},\nonumber\\
\tilde{E}_T(x,0,t)^{S}&=& 0,
\label{gpds_def}
\eea
where
\be
\tau= \frac{\sqrt{t_0- t}}{2M}.
\ee
One cannot  extract the precise form of the
relations between the GPDs and TMDs but one can discuss the
relations of first, second, third, and fourth type, depending
on the number of derivatives of the involved GPDs
in impact parameter space. The relations between these functions have been discussed in detail in Ref. \cite{metz}. 

After obtaining the chiral odd GPDs from helicity amplitudes, we can obtain the
flavor structure of the GPDs using the SU(4) symmetry of the proton
wavefunction as follows \cite{goldstein1}
\bea
F^{u}_T &=& \frac{3}{2} F^{0}_T - \frac{1}{6}F^{1}_T ,\nonumber\\
F^{d}_T &=& - \frac{1}{3}F^{1}_T,
\label{flavor}
\eea
where $F^{q}_T= \{\tilde{H}_T, \tilde{E}_T, H_T, E_T\}$ $(q=u,d).$

\section{GPDs in impact parameter space \label{impact}}
The FT with respect to the transverse momentum transfer $ \Delta_\perp$ gives the GPDs in transverse impact parameter space. We have introduced
$b_{\perp}$ conjugate to $\Delta_\perp$ which gives the distribution of the quarks in the transverse plane. In the present case, we have taken $\zeta=0$ which implies that the momentum transfer is completely in the transverse direction. One can write
\bea
\mathcal{H}(x,b_\perp)&=& \frac{1}{(2\pi)^2} \int d^2 \Delta_\perp
H(x,0,- \Delta_\perp^2) e^{- i  \Delta_\perp \cdot
b_\perp},\nonumber\\
\mathcal{E}(x,b_\perp)&=& \frac{1}{(2\pi)^2} \int d^2 \Delta_\perp
E(x,0,-\Delta_\perp^2) e^{- i \Delta_\perp \cdot
b_\perp}.
\eea
Here $b$=$|b_\perp|$ is the impact parameter measuring the transverse distance
between the struck parton and the center of momentum of the hadron. In the DGLAP
region ($\zeta < x < 1$), \cite{gpds2} the parameter $b$ gives the location of the
quark where it is pulled out and put back to the nucleon whereas in ERBL region ($-\zeta < x < \zeta$)
it describes the location of quark-antiquark pair inside the nucleon.

The fundamental quantity $E_T+2 \tilde{H}_T$ describes the Boer-Mulders function $h_1^{\perp}$ and gives the distribution of polarized quarks inside the unpolarized nucleons in
the opposite direction. The first moment of $E_T(x,0,0)+ 2 \tilde{H}_T(x,0,0)$ is normalized as
\be
\int_{-1}^{1} dx (E_T(x,0,0)+ 2 \tilde{H}_T(x,0,0))= \kappa_T ,
\ee
where $\kappa_T$ gives the average position of quarks considering them in the $\hat{x}- \hat{y}$ plane in such a way that they are with spin along $\hat{x}$ direction and shifted in $\hat{y}$ direction in an unpolarized target relative to the transverse center of momentum. The term $E_T+ 2 \tilde{H}_T$ gives the deformation in the center of momentum frame due to spin-orbit correlation and can be defined in terms of impact parameter space as follows
\be
\mathcal{E}_T(x,b^2)+ 2 \mathcal{\tilde{H}}_T(x,b^2)= \int d^2
 \Delta_\perp e^{-i  b \cdot  \Delta}(E_T(x,0,t)+ 2
\tilde{H}_T(x,0,t)). \label{2h+eimpact}
\ee

From Eqs. (\ref{a0}), (\ref{a1}) and (\ref{gpds_def}) we can compute $E_T+ 2 \tilde{H}_T$ for for the cases of scalar diquark spin-0 and axial-vector spin-1  and the results for $S=0$ and $S=1$ components can be expressed as
\bea
(2 \tilde{H}_T(x,0,t)+E_T(x,0,t))^{0} &=& 2 M \int d^2 k_\perp \frac{g_s^2(1-x)^5 (M x+m)^2}{x \ L_1^2 L_2^2}, \nonumber\\
(2 \tilde{H}_T(x,0,t)+E_T(x,0,t))^{1} &=& -2 M \int d^2 k_\perp \frac{g_A^2 (1-x)^4}{x \ L_1^2 L_2^2},
\label{2h+e}
\eea
where
\bea
L_1=k_\perp^2 - M^2 x (1-x) + M_x^2 x +M_\Lambda^2 (1-x), \nonumber\\
L_2= k_\perp - (1-x) \Delta_\perp^2 -  M^2 x (1-x) + M_x^2 x
+M_\Lambda^2 (1-x).
\eea

\begin{figure}
\minipage{0.42\textwidth}
 \includegraphics[width=8cm]{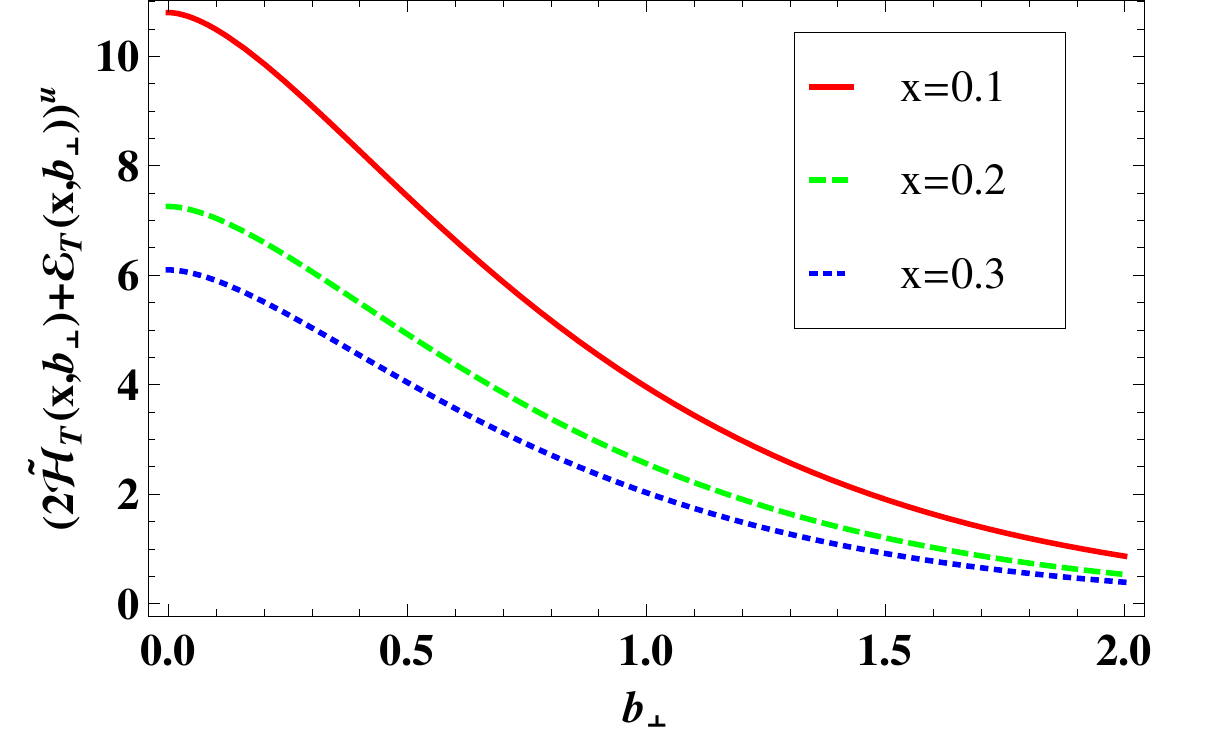}
 \endminipage\hfill
 \minipage{0.42\textwidth}
 \includegraphics[width=8cm]{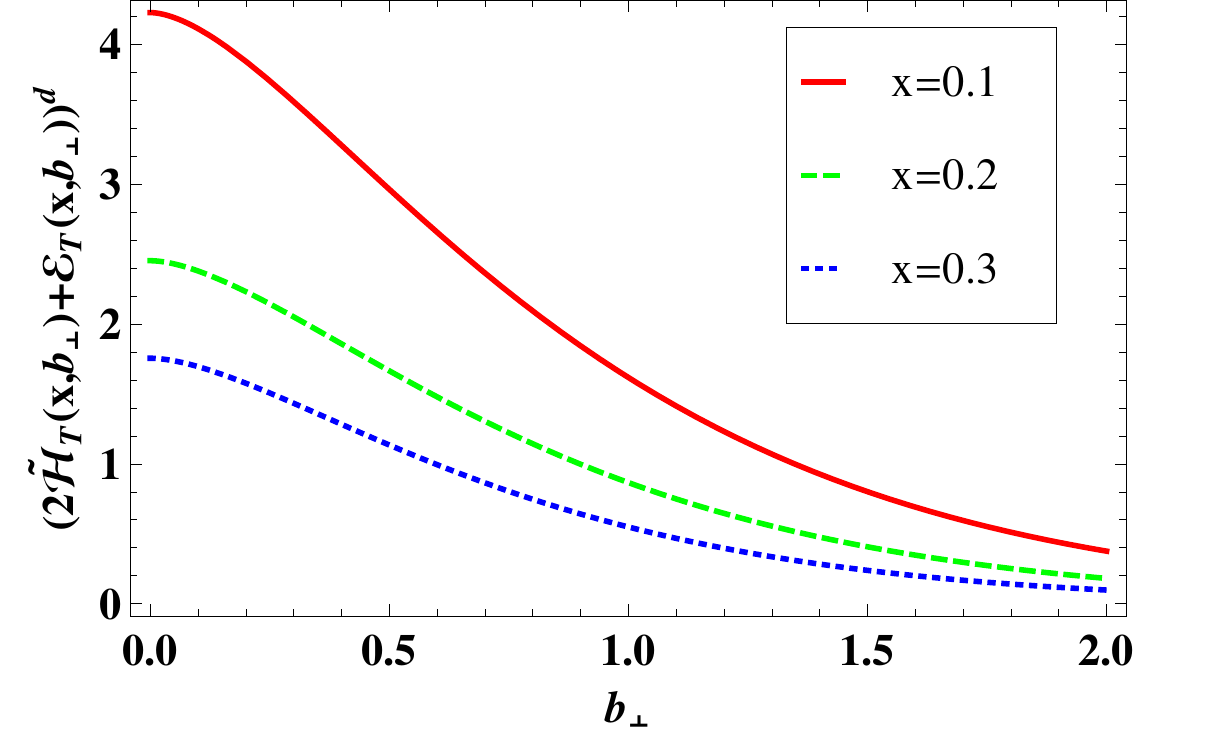}
 \endminipage\hfill
\caption{$\left(\mathcal{E}_T+2 \tilde{\mathcal{H}_T}\right)^q$ ($q=u,d$) as
function of $b_\perp$ for fixed values of $x$ for the up (left panel) and down (right panel) quarks.}
\label{boer-mulders}
\end{figure}

Using Eq. (\ref{flavor}), we can now calculate the explicit contribution for the up and down quarks for each of the GPD and substituting them further in Eq. (\ref{2h+e}) gives the combination $(2 \tilde{H}_T(x,0,t)+E_T(x,0,t))^{S}$. The term $\mathcal{E}_T(x,b^2)+ 2 \mathcal{\tilde{H}}_T(x,b^2)$ in impact parameter space can be obtained from Eq. (\ref{2h+eimpact}).
We have taken the following vale of the masses as input parameters
\bea
M=0.938 {\rm GeV} &,& m=0.5 {\rm GeV}, \nonumber \\
M_x^u=0.4972 {\rm GeV} &,& M_\Lambda^u=0.9728  {\rm GeV}, \nonumber \\
M_x^d=0.7918 {\rm GeV} &,& M_\Lambda^d=0.9214 {\rm GeV}. \eea
We have presented in Fig. \ref{boer-mulders}, $2 \tilde{\mathcal{H}}_T+\mathcal{E}_T$ for the up and down quarks as function of $b_\perp$ for fixed values of $x$.
The magnitude of the term $2\tilde{\mathcal{H}}_T+\mathcal{E}_T$ is found to increase as the
value of the $b_\perp$ decreases or in other words the distribution peaks are highest at
$b_\perp=0$. As the term basically describes the correlation between the quark spin and angular momenta, we can say that the partons are distributed mostly near $b_\perp=0$ which is the center of momentum. As we move away from the center of momentum towards larger values of $b_\perp$, the density of partons decreases. It is also observed that as the value of $x$ increases the
magnitude of $2\tilde{\mathcal{H}}_T+\mathcal{E}_T$ decreases. The difference between the magnitudes of $2\tilde{\mathcal{H}}_T+\mathcal{E}_T$ is more towards the lower values of $b_\perp$.

\section{spin densities \label{spin}}
In this section, we define  the three-dimensional densities: (a) $\rho(x,b_\perp,\lambda,\Lambda)$ which gives the probability to find a quark with momentum fraction $x$ and transverse position $b_\perp$ with light-cone helicity $\lambda (= \pm1)$ and longitudinal polarization $\Lambda (=\pm 1)$. (b) $\rho(x,b_\perp,s_\perp,S_\perp)$ which gives the probability to find a quark with momentum fraction $x$ with transverse position $b_\perp$ and transverse spin $s_\perp$ in the proton with transverse spin $S_\perp$. We have
\be
\rho(x, b_\perp, \lambda, \Lambda)=\frac{1}{2} [ \mathcal{H}(x, {b^2})+ b^j \epsilon^{ji} \frac{S^i}{M} \mathcal{E}'(x,{b^2})+ \lambda \Lambda \mathcal{\tilde{H}}(x, {b^2})],
\label{longitudinal_spin}
\ee
\bea
\rho(x, b_\perp,s_\perp, S_\perp)&=& \frac{1}{2}[\mathcal{H}(x, {b^2})+ s^i S^i \Big(\mathcal{H}_T(x, {b^2})- \frac{1}{4 M^2} \Delta_b \mathcal{\tilde{H}}_T(x, {b^2})\Big)+\frac{b^j \epsilon^{j i}}{M}\Big( S^i \mathcal{E}'(x,{b^2})+ \nonumber\\
&& s^i [\mathcal{E}'_T(x,{b^2})+ 2 \mathcal{\tilde{H'}}_T(x,{b^2})] \Big)+ s^i (2 b^i b^j - b^2 \delta_{i j}) \frac{S^j}{M^2} \mathcal{\tilde{H''}}_T(x,{b^2})].
\label{transverse_spin}
\eea
They depend on $b_\perp$ only via $b_\perp^2=b^2$ due to rotational invariance and we can define
\be
f'= \frac{\partial}{\partial b^2} f, f''= \Big(\frac{\partial}{\partial b^2}\Big)^2 f, \Delta_b f= 4 \frac{\partial}{\partial b^2}\Big(b^2 \frac{\partial}{\partial b^2}\Big)f,
\ee
with two dimensional antisymmetric tensor $\epsilon^{ij}$, $\epsilon^{12}=-\epsilon^{21}=1$ and $\epsilon^{11}=\epsilon^{22}=0$.
TMDs and impact parameter  dependent parton distribution functions (ipdpdfs) are related to each
other and contain valuable information about the structure of the
nucleon \cite{metz}. The relations read as follows
\be
f_1\longleftrightarrow \mathcal{H}, \  f_{1 T}^\perp \longleftrightarrow - \mathcal{E}', \ g_1 \longleftrightarrow \mathcal{\tilde{H}}, \\
h_1 \longleftrightarrow \mathcal{H}_T - \Delta_b \frac{\mathcal{\tilde{H}}_T}{4 M^2}, \ h_1^{\perp} \longleftrightarrow - (\mathcal{E}_T' + 2 \mathcal{\tilde{H}}'_T), \\
h_{1 T}^{\perp} \longleftrightarrow 2 \mathcal{\tilde{H}}''_T,
\label{gpdsandtmds}
\ee
where $f_{1T}^{\perp}$ and $h_{1}^\perp$ are Sivers and Boer-Mulders
distribution functions respectively, $f_1$ denotes the unpolarized quark
distribution, $g_1$ the quark helicity distribution and $h_1$ is the quark
transversity distribution.

In order to study the spin densities in the present
model for the up and down quarks, we have fixed the value of $x$ here.
In Eq. (\ref{longitudinal_spin}), the first term
$\mathcal{H}$ describes the density of unpolarized quarks in the
unpolarized proton. The term with $\mathcal{\tilde{H}}$ reflects the difference
in density of quarks with helicity either being equal or opposite to the proton helicity.
The term containing $\mathcal{E}'$ describes a sideways shift in the unpolarized parton density
when the proton is transversely polarized.
Eq. (\ref{transverse_spin}) receives contribution from the monopole $\frac{\mathcal{H}}{2}$, dipole $\frac{-1}{2} s_i b_j (\mathcal{E'}_T+2
\tilde{\mathcal{H'}}_T/ M$ and quadrupole $\frac{1}{2} s_i S_i (b_x^2 -
b_y^2) \tilde{\mathcal{H''}}_T/M^2$ terms. The monopole term $\frac{\mathcal{H}}{2}$ in Eq. (\ref{longitudinal_spin}) describing the
unpolarized quark density  gets further modified due to the chiral odd terms $\mathcal{H}_T$ and $\Delta_b \mathcal{\tilde{H}}_T$ in Eq. (\ref{transverse_spin}) where both quark and the proton are transversely polarized. The dipole structure can
be either obtained from the chiral even $\mathcal{E}'$ appearing in the longitudinal
spin distribution (Eq. (\ref{longitudinal_spin})), from the chiral odd contribution $\mathcal{E}'_T + 2
\mathcal{\tilde{H}}'_T$ from the transversely polarized quarks in (Eq. (\ref{transverse_spin})) or both. The term
$\mathcal{\tilde{H}}''_T$  in Eq. (\ref{transverse_spin}) describes the quadrupole structure when both quark
and proton are transversely polarized.

The chiral even terms
$H(x,0,t)$ and $E(x,0,t)$ can be obtained following Ref. \cite{goldstein} and the expression for S=0 and S=1 diquark are respectively expressed as
\bea
H^{0}(x,0,t)&=&\frac{1}{(1-x)} \left( (M x+m)^2 (1-x)^4 I_3 + (1-x)^4
\left(\frac{I_1+I_2}{2}+  (M^2 x (1-x)- M_x^2 x -\right.\right. \nonumber\\
&&\left.\left. M_\Lambda^2(1-x)-m^2 (1-x)- \frac{(1-x)^2 \Delta_\perp^2}{2})I_3\right)\right), \nonumber\\
H^{1}(x,0,t)&=&-\frac{2 (1-x)^2}{x}\left(\frac{I_1+I_2}{2}+I_3\left(M^2 x
(1-x)- M_x^2 x- M_\Lambda^2 (1-x)- \frac{(1-x)^2 \Delta_\perp^2}{2}\right)\right),\nonumber\\
E^{0}(x,0,t)&=& - 2 M (M x+m)(1-x)^4 I_3, \nonumber\\
E^{1}(x,0,t)&=& 2 M (M x+m) (1-x)^4 I_3.
\eea
Here \bea
I_1 &=& \pi \int_{0}^{1} \frac{(1-\alpha) d\alpha}{D^2},\nonumber \\
 I_2 &=& \pi \int_{0}^{1} \frac{\alpha d\alpha}{D^2},\nonumber \\ I_3&=&\pi \int_{0}^{1} \frac{\alpha (1-\alpha) d\alpha}{D^3}, \nonumber\\
D&=& \alpha (1-\alpha) (1-x)^2 \Delta_\perp^2 - M^2 x (1-x) + M_x^2 x+ M_\Lambda^2 (1-x).
\eea

The FT for different contributions discussed in Eqs. (\ref{longitudinal_spin}) and (\ref{transverse_spin}) are expressed as
\be
-\frac{1}{2} \frac{s_i b_j (\mathcal{E}_T'(x,b_\perp) \ + 2 \mathcal{\tilde{H}}'_T(x,b_\perp))}{M}= \frac{s_i b_j}{2 M} \int \frac{\Delta^2 d\Delta}{2 \pi} J_1(\Delta b) (E_T (x,0,t) \ + 2 \tilde{H}_T(x,0,t)),
\ee
\bea
\frac{1}{2} s_i S_i \Big(\mathcal{H}_T(x,b_\perp) - \Delta_b \frac{\mathcal{\tilde{H}}_T(x,b_\perp)}{4 M^2}\Big)&=& \frac{1}{2} s_i S_i \Big( \int \frac{\Delta d\Delta}{2 \pi} J_0(\Delta b) H_T(x,0,t)+ \frac{4}{4 M^2} \int \frac{\Delta^2 d\Delta}{2 \pi} \times \nonumber\\
&&\Big(J_1(\Delta b) + \frac{1}{2} \Delta b  (J_0(\Delta b)- J_2(\Delta b)\Big)\tilde{H}_T(x,0,t) \Big),
\eea
\be
\frac{1}{2} s_i S_i (b_i^2-b_j^2) \frac{\mathcal{\tilde{H}}''_T(x,b_\perp)}{M^2}= \frac{1}{2 M^2} s_i S_i (b_i^2-b_j^2) \int \frac{\Delta d\Delta}{2 \pi} \Big( - \frac{1}{2} \Delta^2 J_0(\Delta b) \tilde{H}_T(x,0,t) \Big), 
\ee
\bea
-\frac{1}{2} S_i b_j \frac{\mathcal{E}'(x,b_\perp)}{M}&=& \frac{1}{4 \pi M} S_i
b_j \int \Delta^2 J_1(\Delta b) E(x,0,t) d\Delta, \\
\frac{1}{2} S_j b_i \mathcal{E}'(x,b_\perp)&=& - \frac{1}{4 \pi} S_j b_i \int
\Delta^2 J_1(\Delta b) E(x,0,t) d\Delta.
\eea
\[\frac{1}{2} \Big(\frac{S_j b_i \mathcal{E}'(x,b_\perp)- s_i b_j\left(\mathcal{E}_T'(x,b_\perp) \ + 2
\mathcal{\tilde{H}}'_T(x,b_\perp)\right)}{M} \Big)= -\frac{1}{4 \pi} S_j b_i \int \Delta^2 J_1(\Delta
b) E(x,0,t) d\Delta \]\be + \frac{1}{4 \pi}  \frac{S_i b_j}{M} \int \Delta^2 d\Delta J_1(\Delta b) (E_T'(x,0,t)+ 2 \tilde{H}'_T(x,0,t))
\ee
where $J_0(\Delta b), J_1(\Delta b), J_2(\Delta b)$ are the Bessel functions of
first kind.

\begin{figure}[!]
\minipage{0.42\textwidth}
   \includegraphics[width=10cm,angle=0]{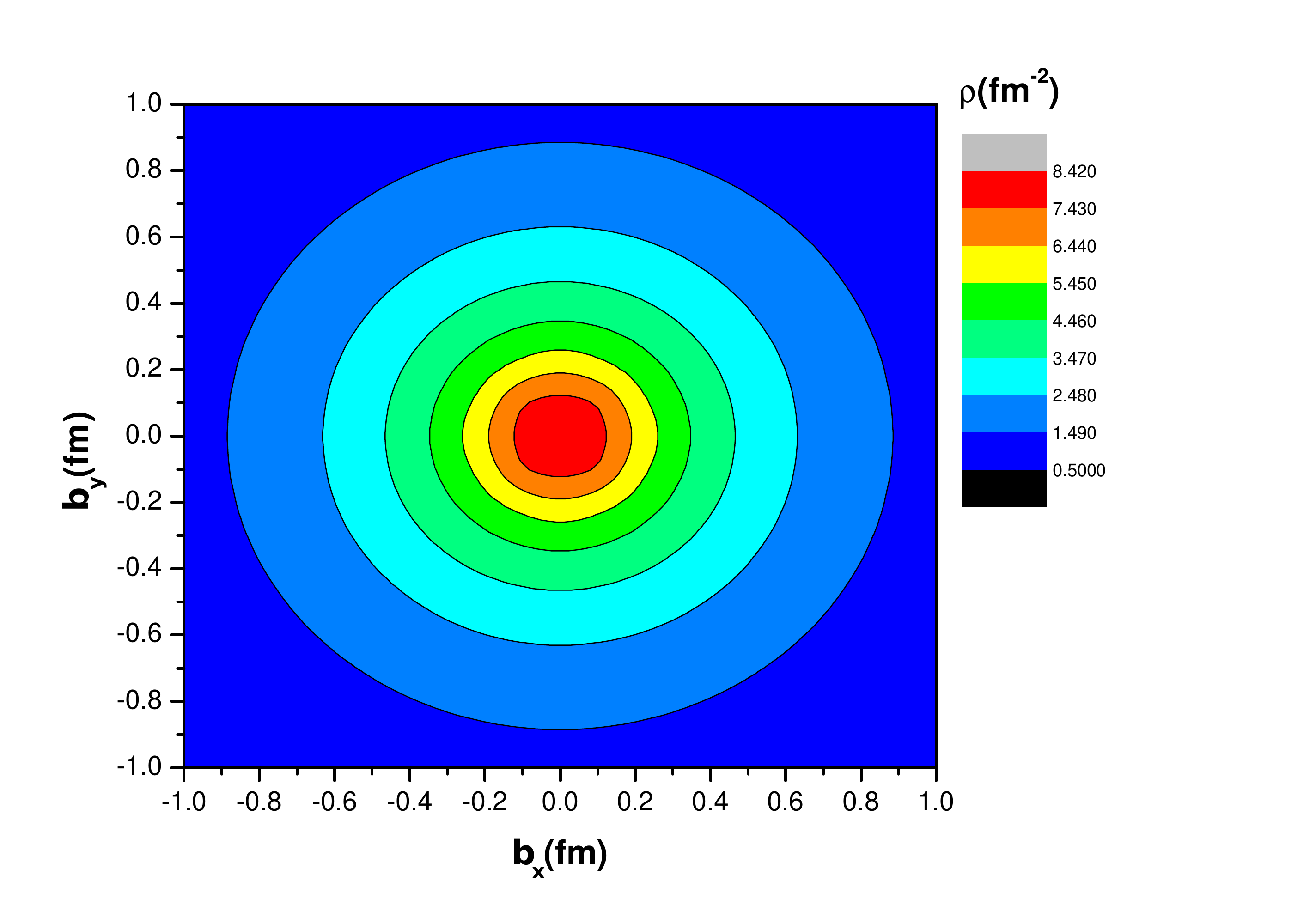}
  \endminipage\hfill
  \minipage{0.42\textwidth}
 \includegraphics[width=10cm,angle=0]{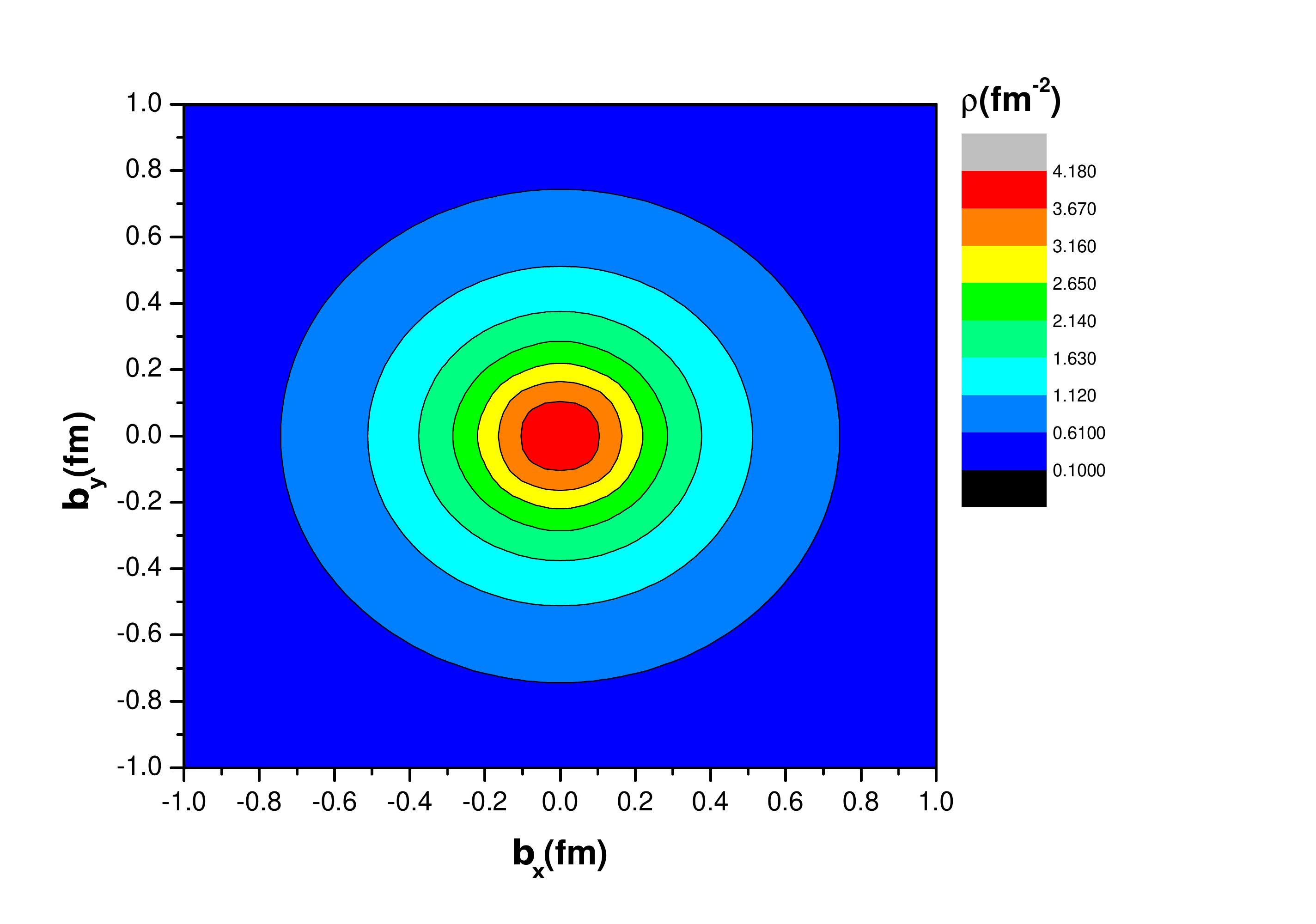}
  \endminipage\hfill
    \caption{The monopole contribution $\mathcal{H}/2$ for the unpolarized quarks
in the unpolarized proton for the up (left panel) and down (right panel)
quarks.}
\label{monopole}
\end{figure}

\begin{figure}[!]
\minipage{0.42\textwidth}
   \includegraphics[width=10cm,angle=0]{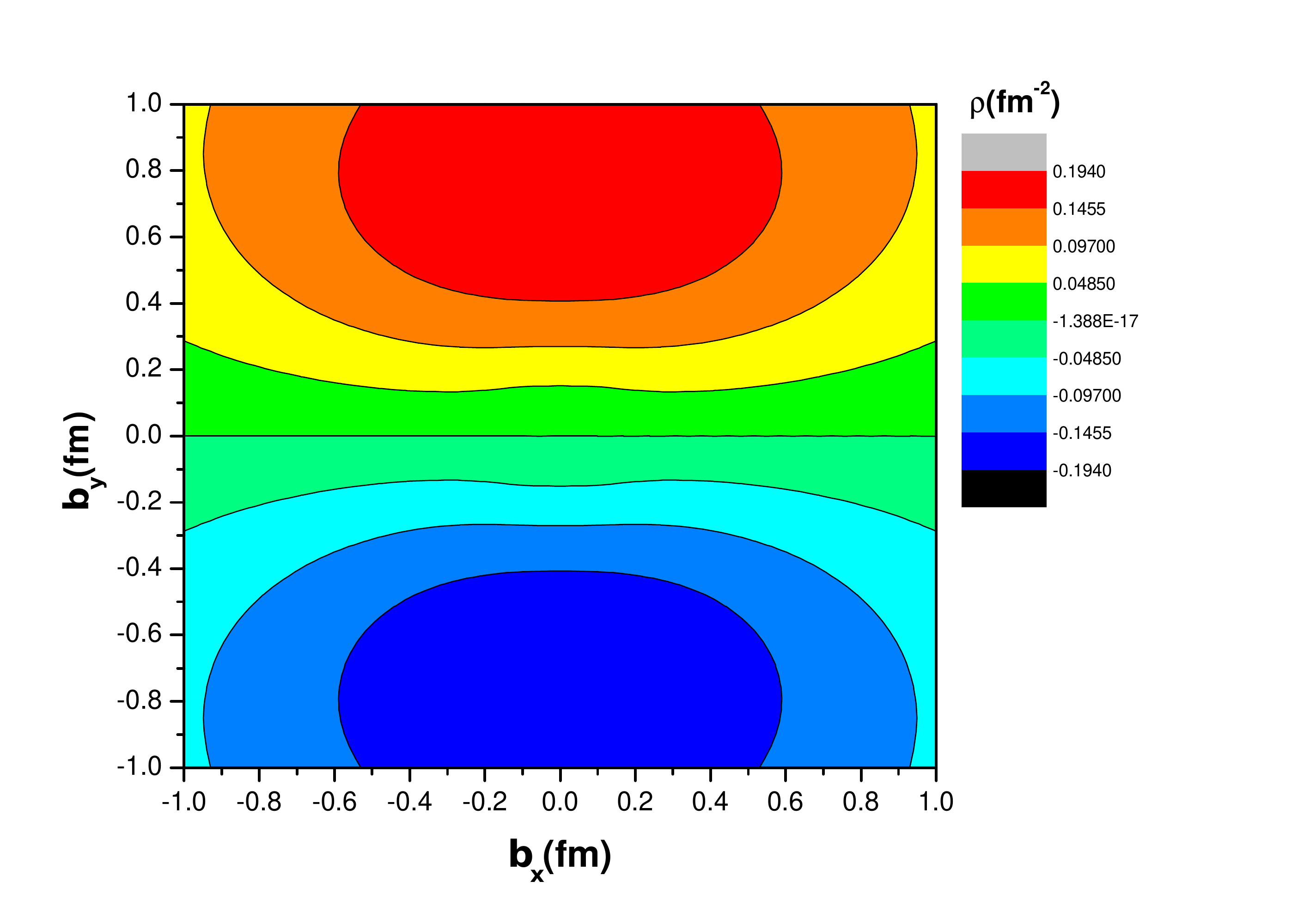}
  \endminipage\hfill
  \minipage{0.42\textwidth}
 \includegraphics[width=10cm,angle=0]{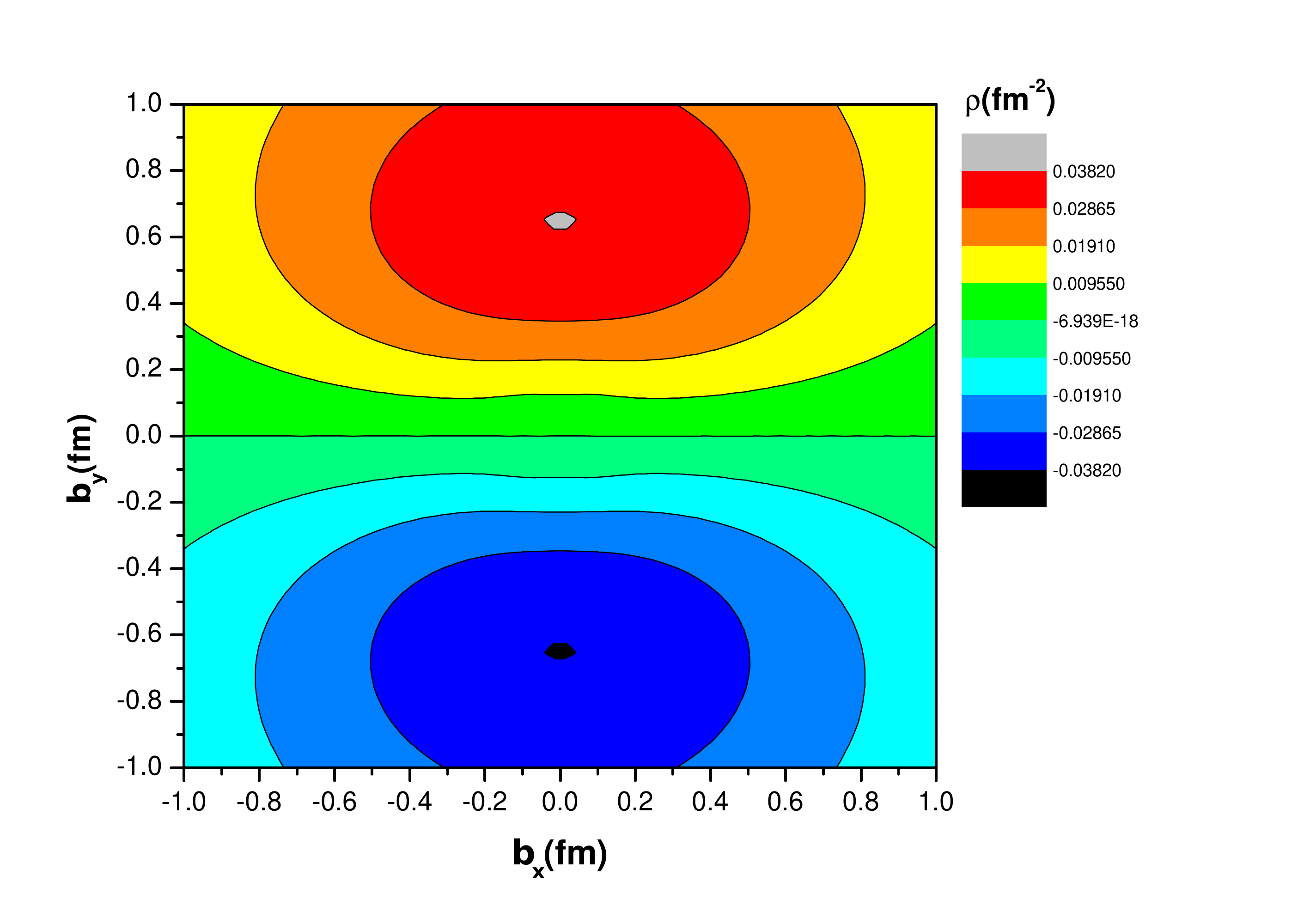}
  \endminipage\hfill
   \caption{The dipole contribution $-\frac{1}{2} s_i b_j (\mathcal{E}'_T + 2
\tilde{\mathcal{H}}'_T )/M$ for the transversely polarized quarks in the
unpolarized proton for the up (left panel) and down (right panel) quarks. }
\label{et+2ht_spin_density}
\end{figure}

\begin{figure}[!]
\minipage{0.42\textwidth}
   \includegraphics[width=10cm,angle=0]{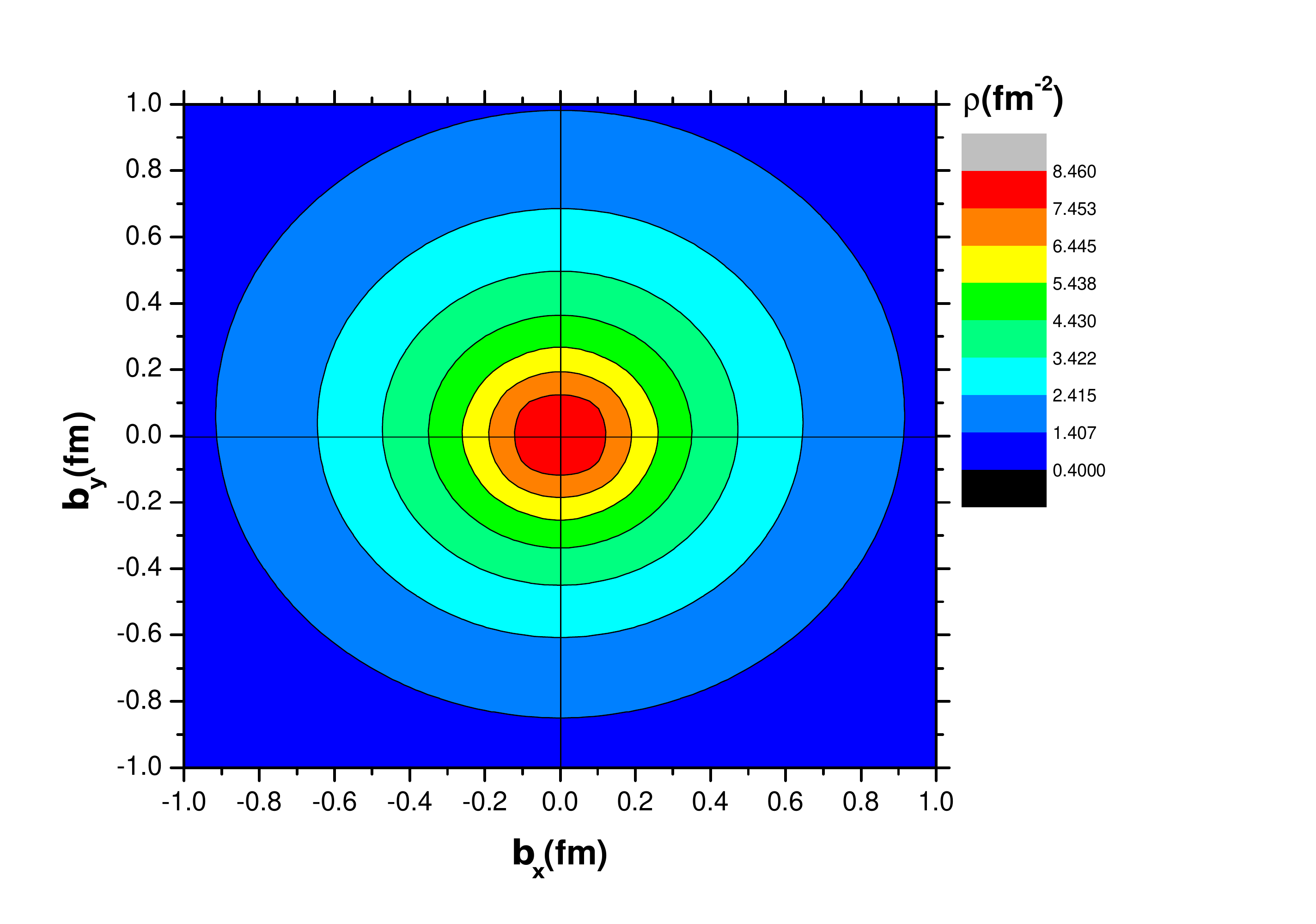}
  \endminipage\hfill
  \minipage{0.42\textwidth}
 \includegraphics[width=10cm,angle=0]{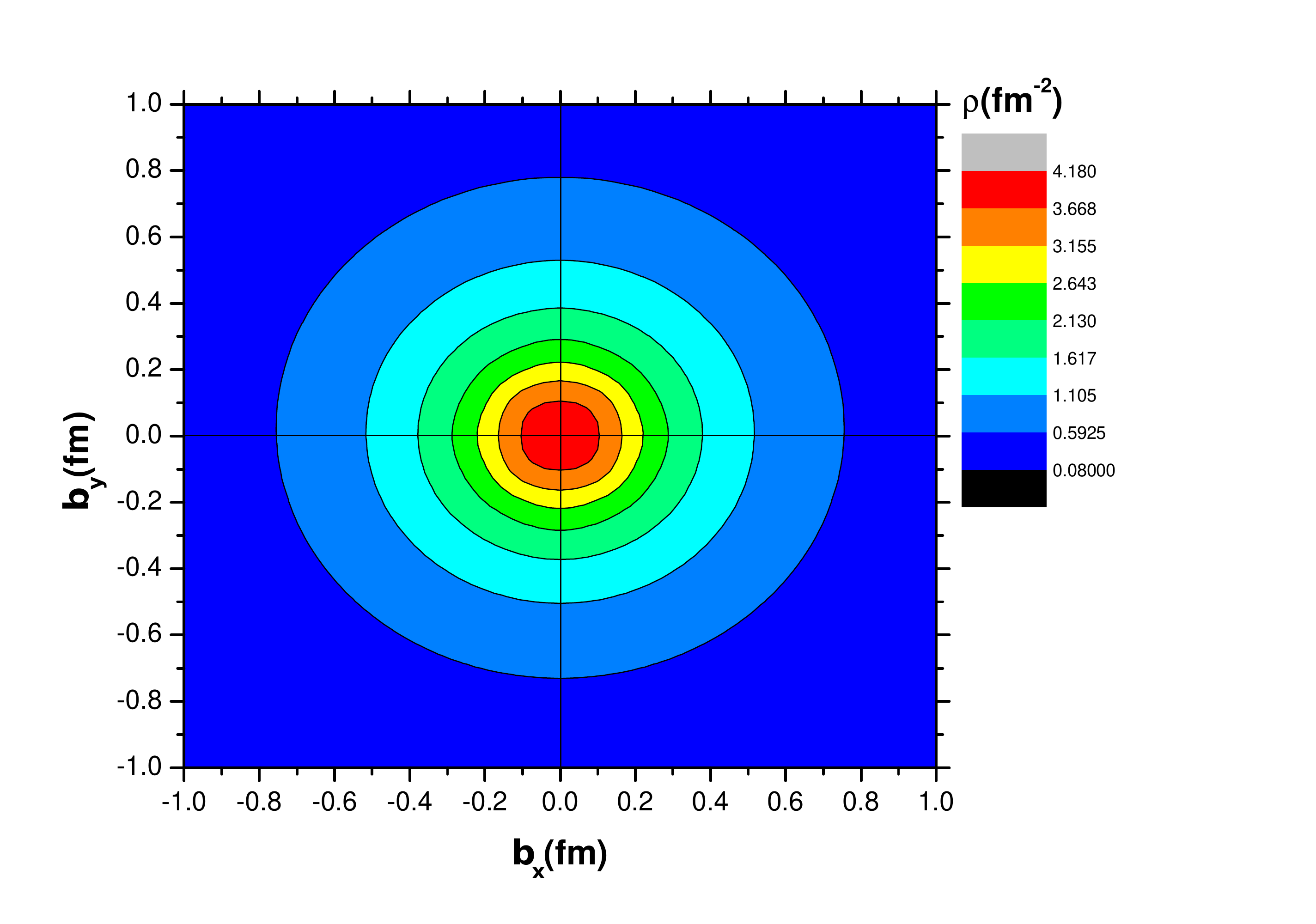}
  \endminipage\hfill
\caption{The sum of monopole  $\mathcal{H}/2$  and dipole contribution
$-\frac{1}{2} s_i b_j (\mathcal{E'}_T + 2 \tilde{\mathcal{H}}'_T )$ for
the up (left panel) and down (right panel) quarks.}
\label{chiral_even_H_2ht+et}
\end{figure}

As, in the present work, we are emphasizing on the spin densities of the valence quarks in impact parameter space, the
value of $x$ is taken to be fixed as $x=0.5$. This is primarily because the valence quarks are supposed to dominate at large and intermediate $x$
$(x \ge 0.2)$. To get the clear picture of the densities of various
contributions we have plotted them as function of $b_x$ and $b_y$ at fixed values of $x$. In Fig.
\ref{monopole}, we have presented the result for the monopole contribution for
unpolarized quarks in an unpolarized proton for the up and down quarks $\mathcal{H}/2$. We observe that the distribution for the up quark is more spread as compared to the distribution of the down quark and is almost twice.  In Fig. \ref{et+2ht_spin_density} we have
presented the results for the dipole contribution $-\frac{1}{2} s_i b_j (\mathcal{E'}_T + 2
\tilde{\mathcal{H}}'_T )/M$ for $\hat{x}$ polarized quarks in
an unpolarized proton. It is observed that the distribution has a reflection symmetry along the $\hat{y}$ direction and all orientations are equally probable in the positive and negative $\hat{y}$ direction. The density obtained for the up quark is however greater than the density obtained for the
down quark. In case the monopole term $\mathcal{H}/2$ is multiplied by the quark charge and
the sum over all the flavours of quarks is taken, the nucleon parton
charge density in the transverse plane is obtained.
In Fig. \ref{chiral_even_H_2ht+et}, we
have shown the results for the sum of the monopole $\mathcal{H}/2$  and the dipole $-\frac{1}{2} s_i b_j (\mathcal{E'}_T + 2 \tilde{\mathcal{H}}'_T )$ contributions. This results in the distortion in the impact parameter space and the distortion is found to be towards the +ve y-axis for the up quark. A comparatively smaller distortion is observed  for the down quark. When the quarks are
transversely polarized in an unpolarized proton, the dipole contribution
introduces a large distortion transverse to both the quark spin and the momentum
of the proton. This in turn suggests that quarks in this situation also have a
transverse component of orbital angular momentum which is related to the
non-vanishing of the Boer-Mulders function describing the distribution of the polarized quarks inside the unpolarized
proton. Thus, a large distortion suggests a large value of first moment of $E_T+2 \tilde{H}_T$.

\begin{figure}[!]
\minipage{0.42\textwidth}
   \includegraphics[width=10cm]{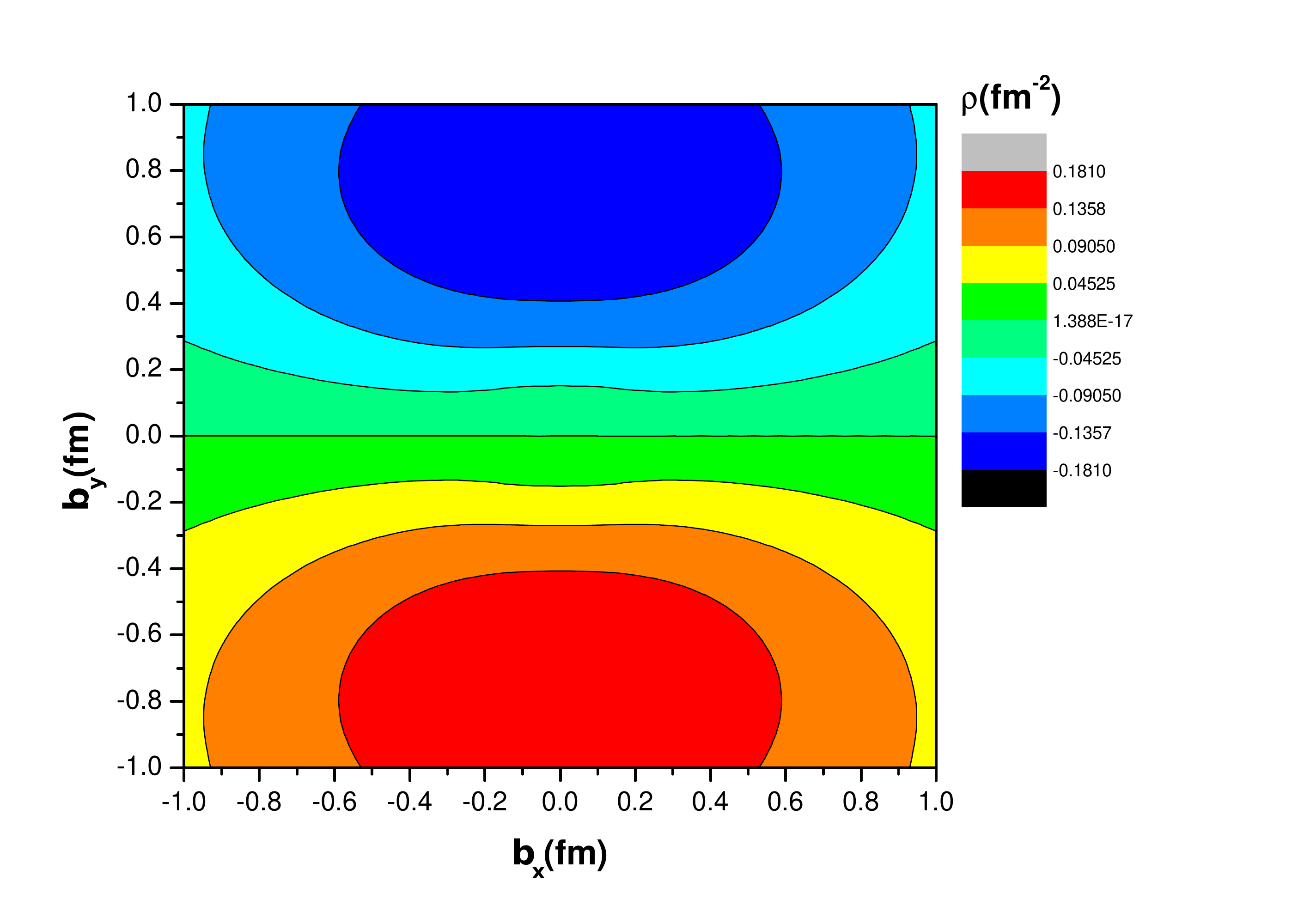}
  \endminipage\hfill
  \minipage{0.42\textwidth}
 \includegraphics[width=10cm]{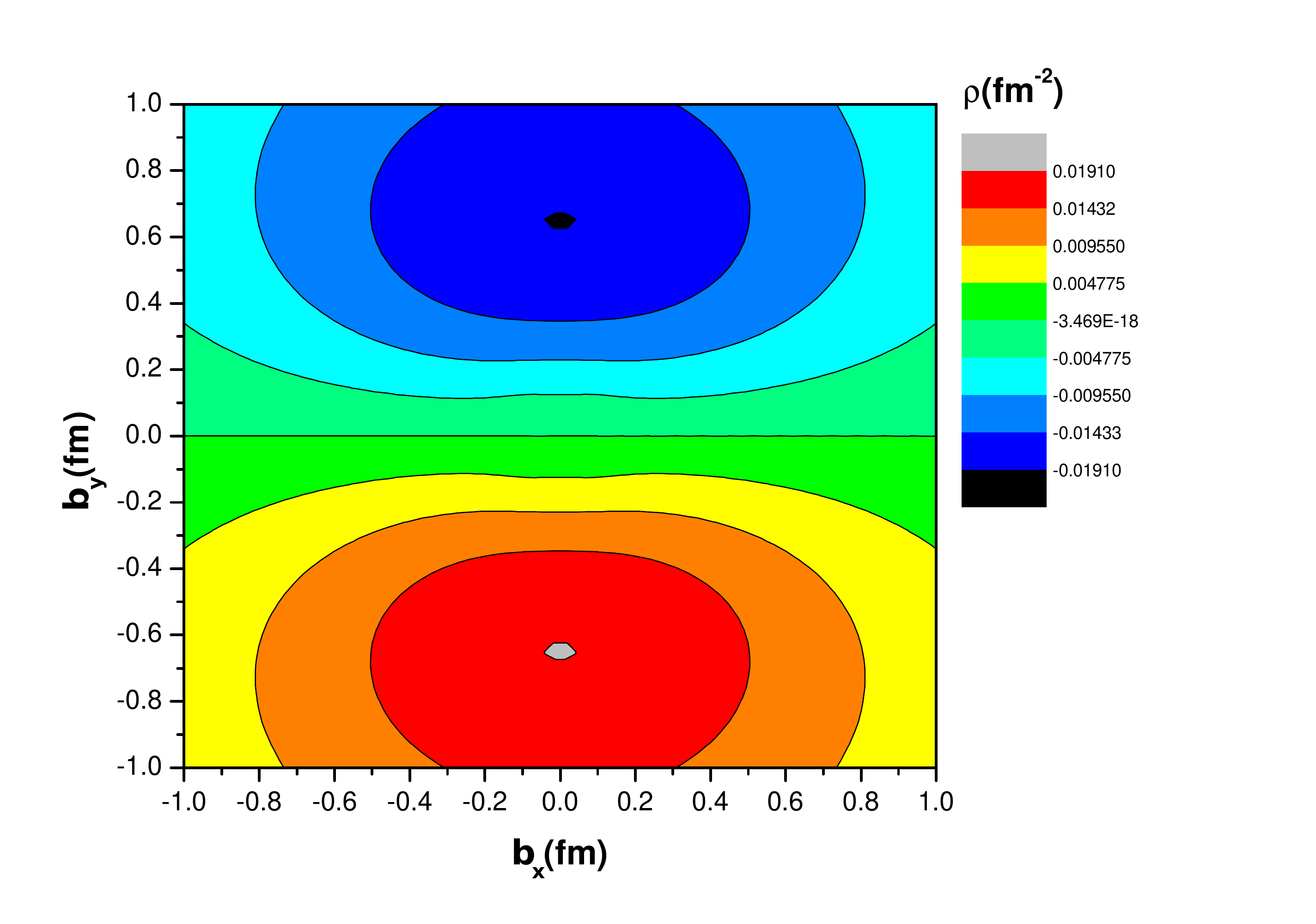}
  \endminipage\hfill
   \caption{The dipole contribution $-\frac{1}{2} S_i b_j \mathcal{E'}$ for the unpolarized
quarks in the transversely polarized proton for the up (left panel) and down (right
panel) quarks.}
\label{dipole_chiral_even_E}
\end{figure}

\begin{figure}[!]
\minipage{0.42\textwidth}
 \includegraphics[width=10cm,angle=0]{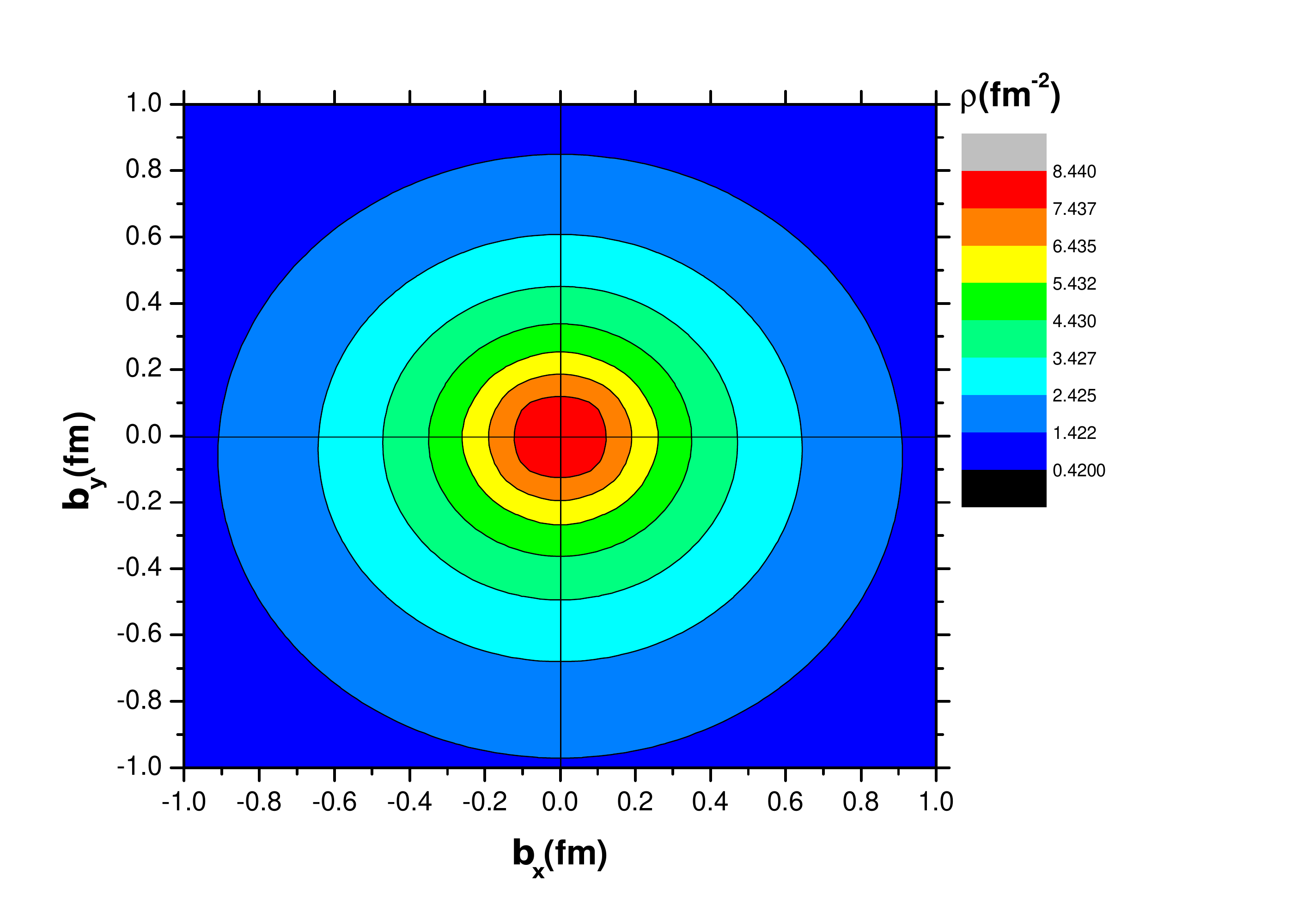}
  \endminipage\hfill
\minipage{0.42\textwidth}
 \includegraphics[width=10cm,angle=0]{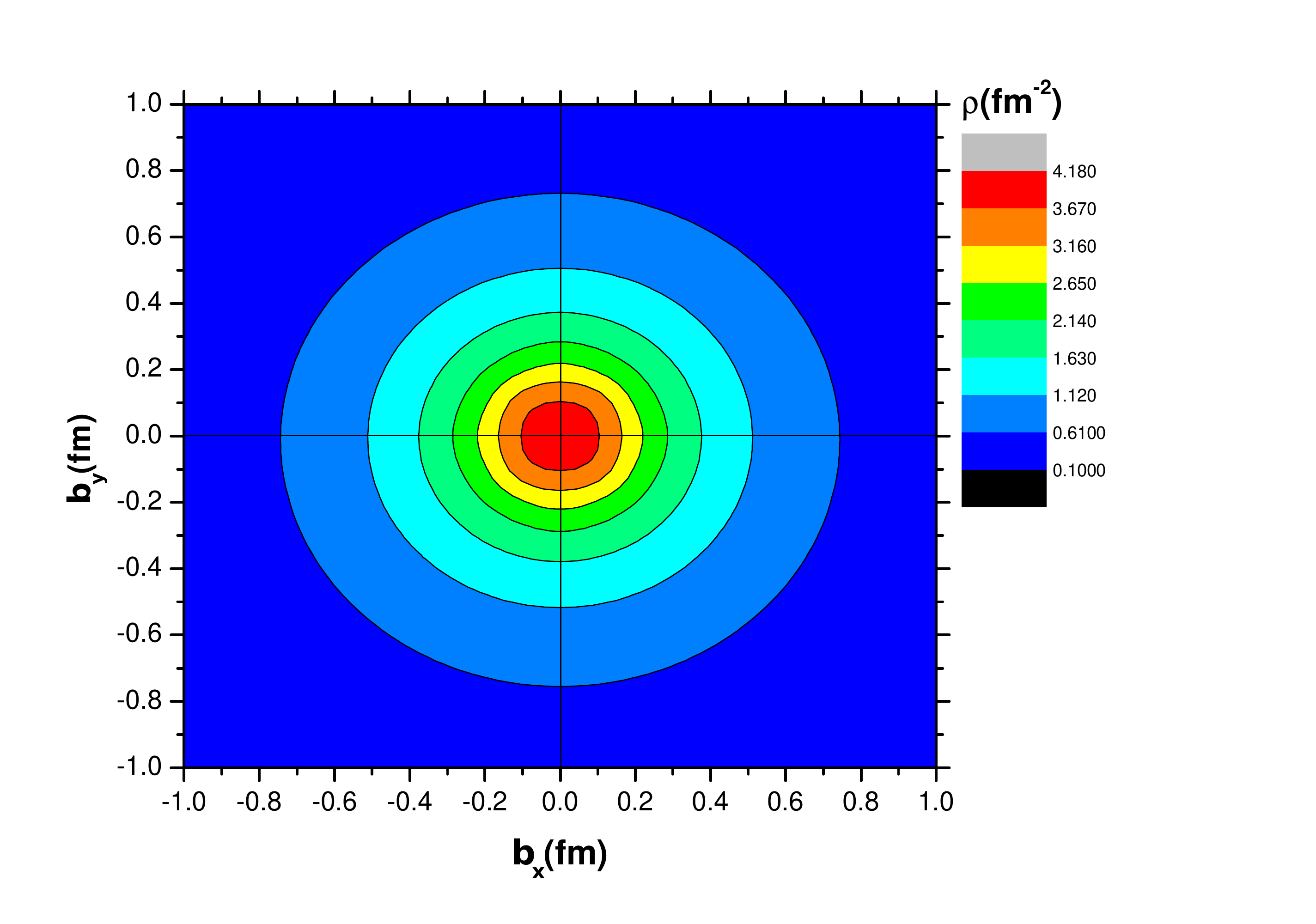}
  \endminipage\hfill
   \caption{The sum of monopole  $\mathcal{H}/2$  and dipole contribution
$-\frac{1}{2} S_i b_j \mathcal{E}'/M$ for the unpolarized quarks in the transversely
proton for the up (left panel) and down (right panel) quarks.}
\label{sum_monopole_dipole_chiral_even_H_chiral_even_E}
\end{figure}

In Fig. \ref{dipole_chiral_even_E} the results of the
dipole contribution $-\frac{1}{2} S_i b_j \mathcal{E'}$ for the unpolarized
quarks in the transversely polarized proton have been presented. It is clear that the
dipole contribution is twice as larger for the up quark as compared to the down quark and the distribution is more
spread over the $b_x$ and $b_y$ plane for the up quark than the down quark. In Fig.
\ref{sum_monopole_dipole_chiral_even_H_chiral_even_E} we have shown  the
results for the sum of contributions coming from the monopole $\mathcal{H}$ and the
dipole term $\frac{-1}{2} S_i b_j \mathcal{E'}$
(which in this case is for transversely $\hat{x}$-polarized proton for the up and
down quarks). One can see that the distortion is obtained and it is larger
for the up quark  than for the down quark. This is basically due to the presence
of the $\mathcal{E}'$ term which already has a large magnitude for the up quark.


\begin{figure}[!]
\minipage{0.42\textwidth}
   \includegraphics[width=10cm]{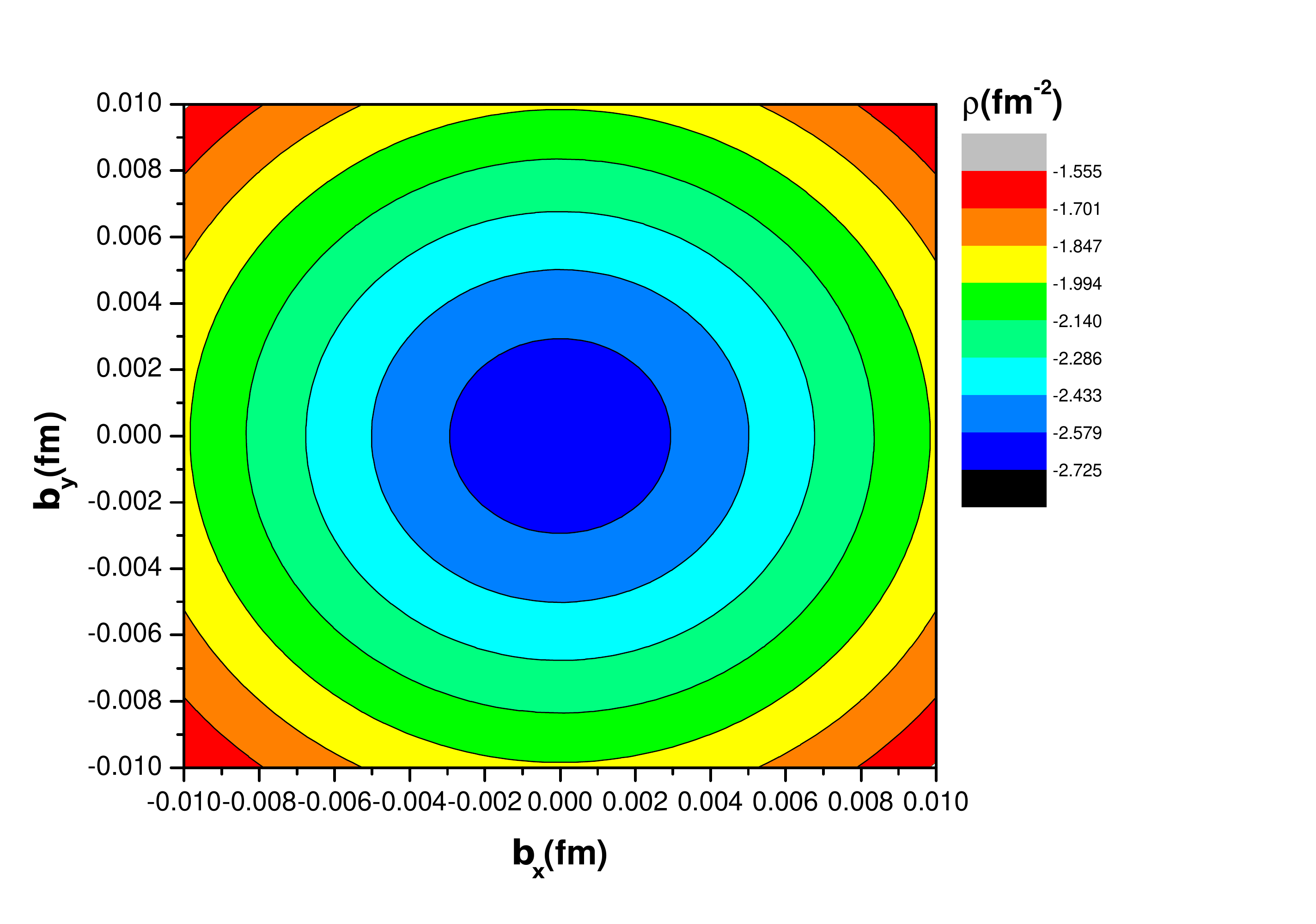}
  \endminipage\hfill
  \minipage{0.42\textwidth}
  \includegraphics[width=10cm]{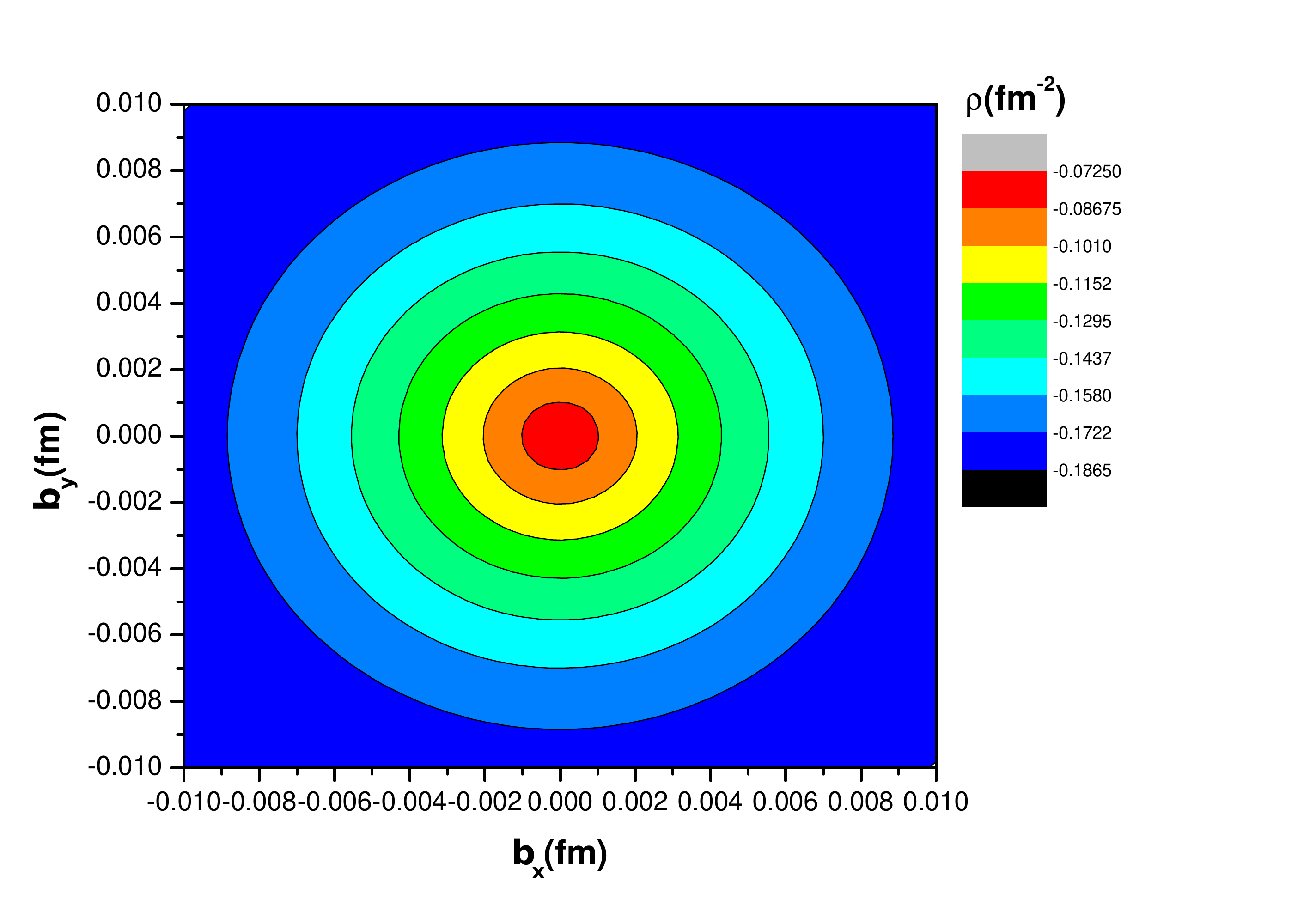}
  \endminipage\hfill
 \caption{The monopole contribution $\frac{1}{2} s_i S_i
(\mathcal{H}_T- \Delta_b \tilde{\mathcal{H}}'_T/{4 M^2})$ for
the quarks in the proton polarized in the same direction for the up (left panel) and
down (right panel) quarks.}
\label{monopole_chiral_odd_terms}
\end{figure}
\begin{figure}[!]
\minipage{0.42\textwidth}
   \includegraphics[width=10cm]{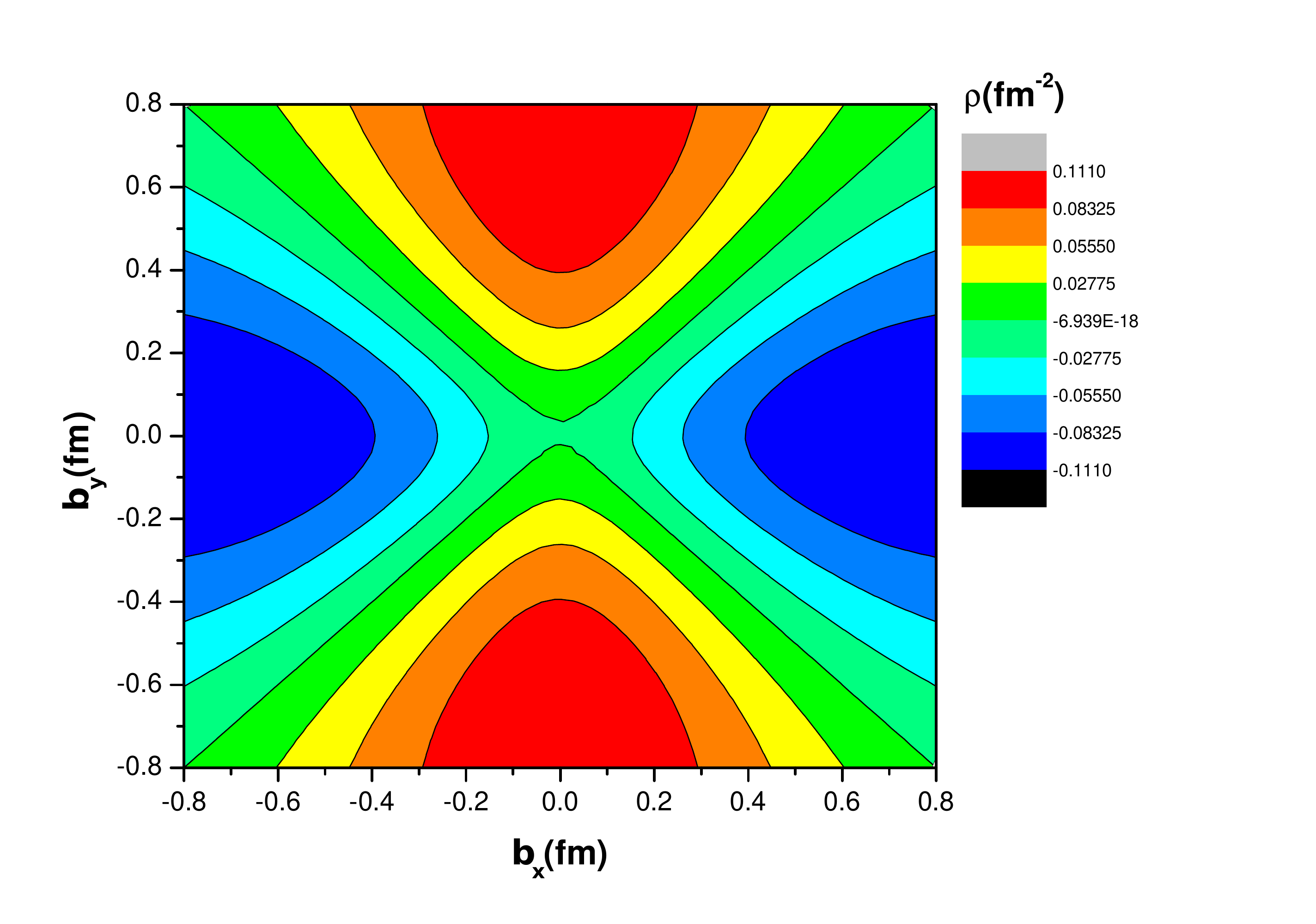}
 \endminipage\hfill
  \minipage{0.42\textwidth}
  \includegraphics[width=10cm]{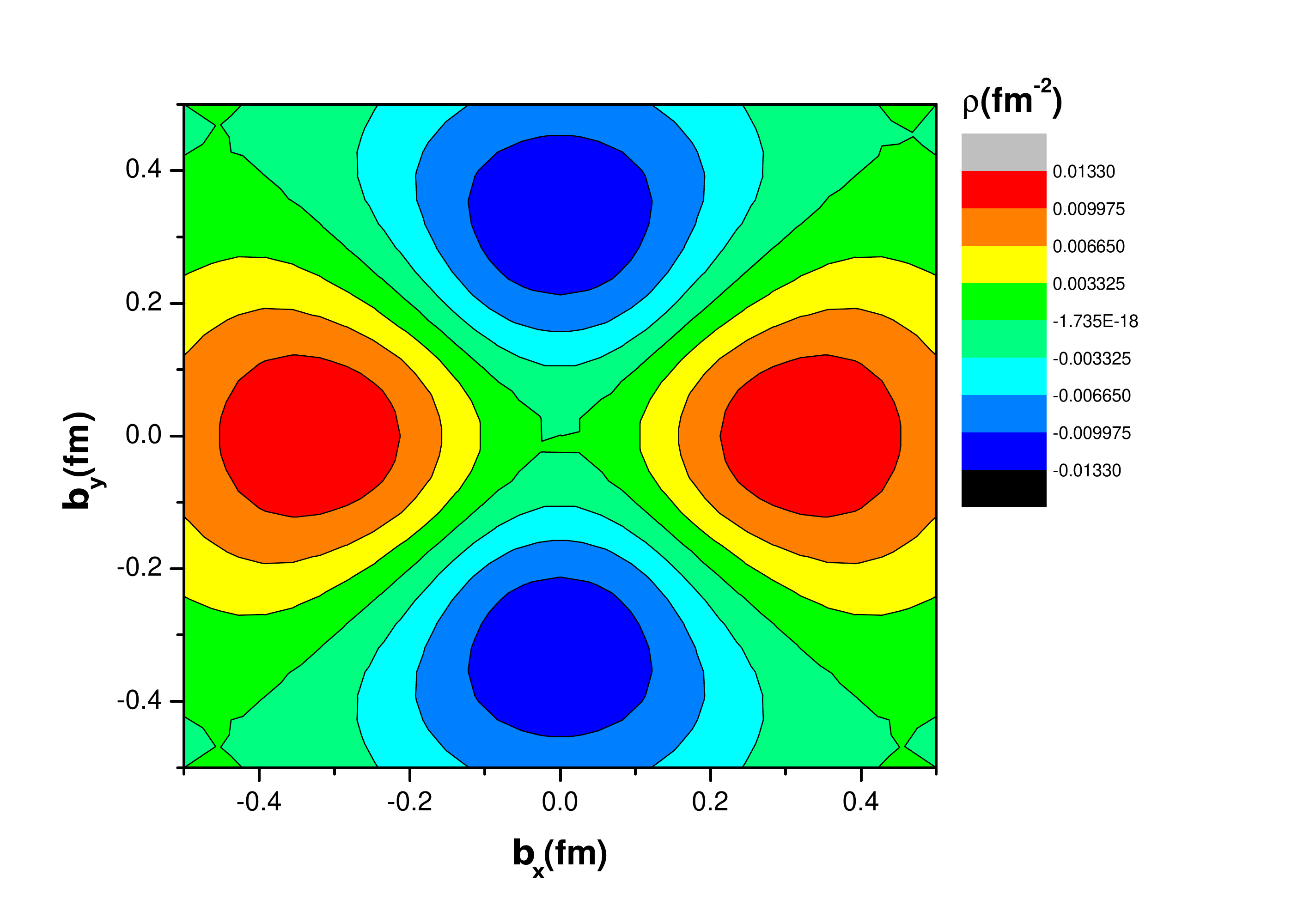}
  \endminipage\hfill
   \caption{The quadrupole contribution $\frac{1}{2} s_i S_i (b_i^2 - b_j^2)
\tilde{\mathcal{H}}''_T/ M^2$ for the
quarks in the proton polarized in the same direction for the up (left panel) and
down (right panel) quarks.}
\label{_quadrupole_chiral_odd_H_double_tilde}
\end{figure}

In Fig. \ref{monopole_chiral_odd_terms} and
\ref{_quadrupole_chiral_odd_H_double_tilde} we have presented the results for the monopole $\frac{1}{2} s_i S_i
(\mathcal{H}_T- \Delta_b \tilde{\mathcal{H}}'_T/{4 M^2})$  and the quadrupole $\frac{1}{2} s_i S_i (b_i^2 - b_j^2)
\tilde{\mathcal{H}}''_T/ M^2$ terms where the
quarks and protons are transversely polarized. We observe that the sign
flips for up and down quarks for both contributions. The opposite sign for
monopole $\frac{1}{2} s_i S_i
(\mathcal{H}_T- \Delta_b \tilde{\mathcal{H}}'_T/{4 M^2})$  and quadrupole $\frac{1}{2} s_i S_i (b_i^2 - b_j^2)
\tilde{\mathcal{H}}''_T/ M^2$ term is due to the sign difference in the up and down quark's
$x$ dependence of $H_T$ and $\tilde{H}'_T$ as predicted by the model. It is also seen that the monopole term for the up quark is more spread than the down
quark and in the case of quadrupole contribution, the up quark distribution is again more spread as compared to the down quark distribution  which is spread over the small region in the plane with opposite sign. Similar results have also been obtained for nucleon spin densities in light-front
constituent quark model \cite{pincetti,pasquini}.

\begin{figure}[!]
\minipage{0.42\textwidth}
   \includegraphics[width=10cm]{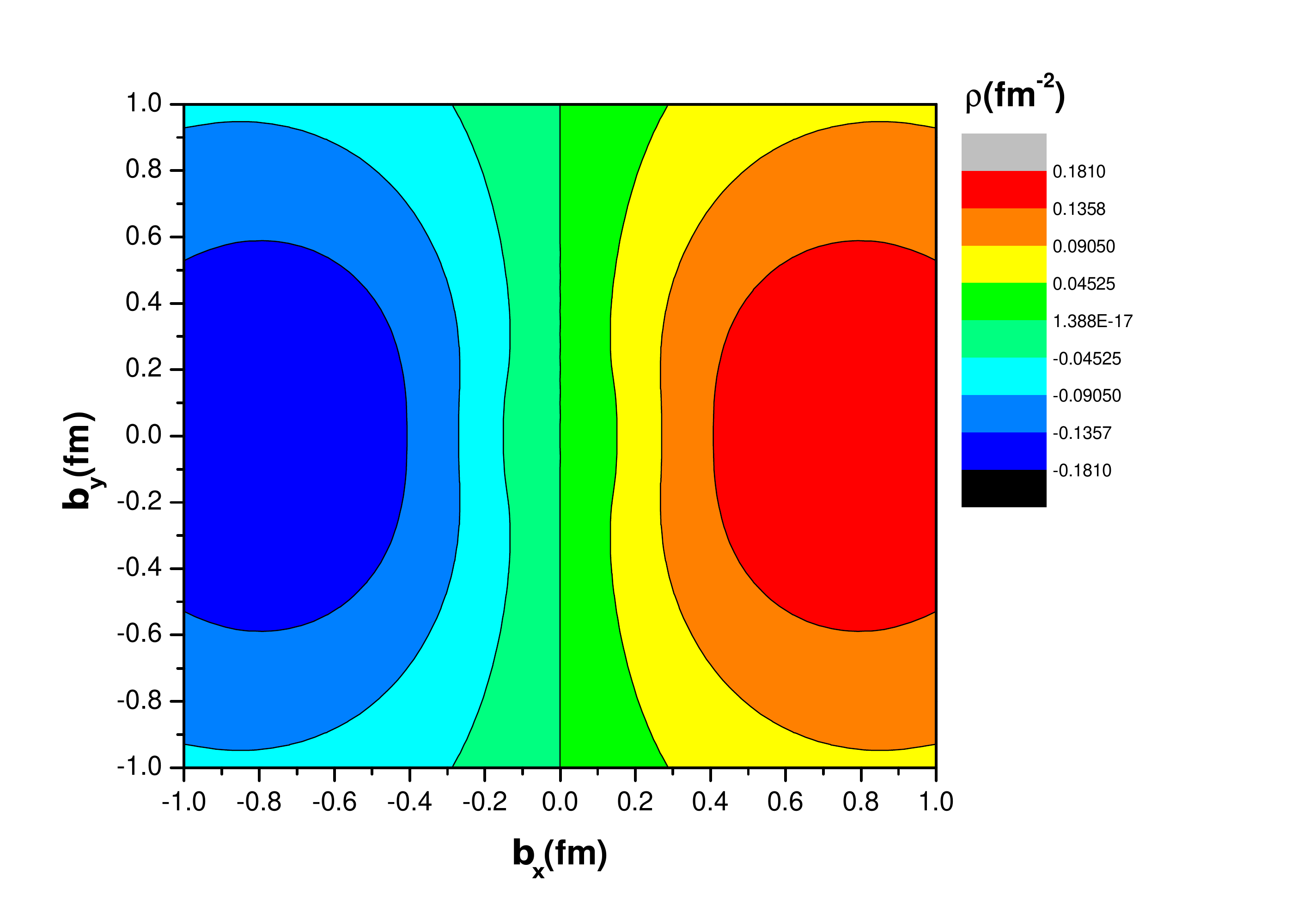}
 \endminipage\hfill
  \minipage{0.42\textwidth}
  \includegraphics[width=10cm]{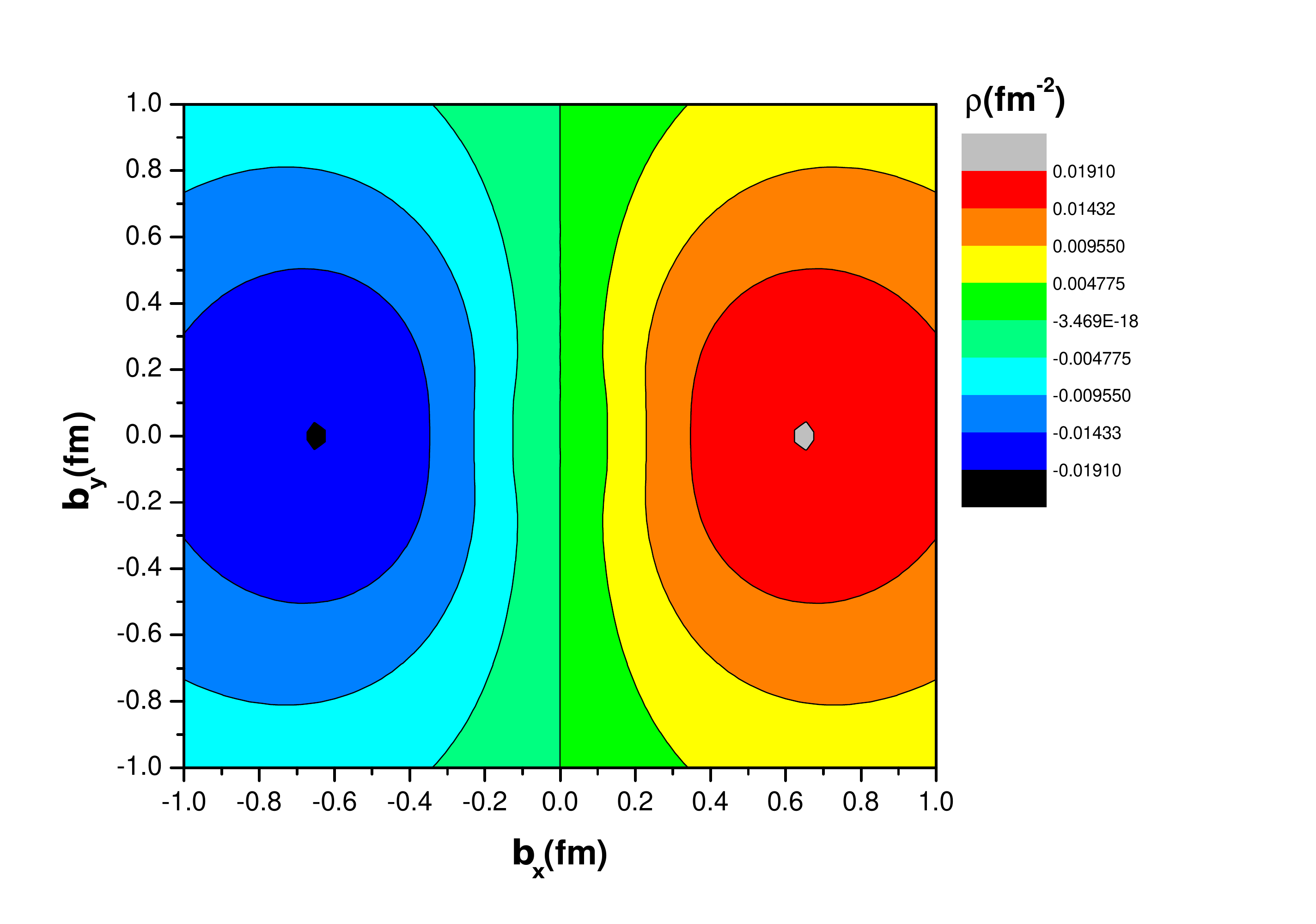}
  \endminipage\hfill
   \caption{The dipole contribution $\frac{1}{2} S_j b_i \mathcal{E'}/M$ for $\hat{x}$ polarized quarks in a proton polarized in $\hat{y}$ direction for the up (left panel) and down (right panel) quarks.}
\label{dipole_chiral_even_x_axis}
\end{figure}

\begin{figure}[!]
\minipage{0.42\textwidth}
   \includegraphics[width=10cm]{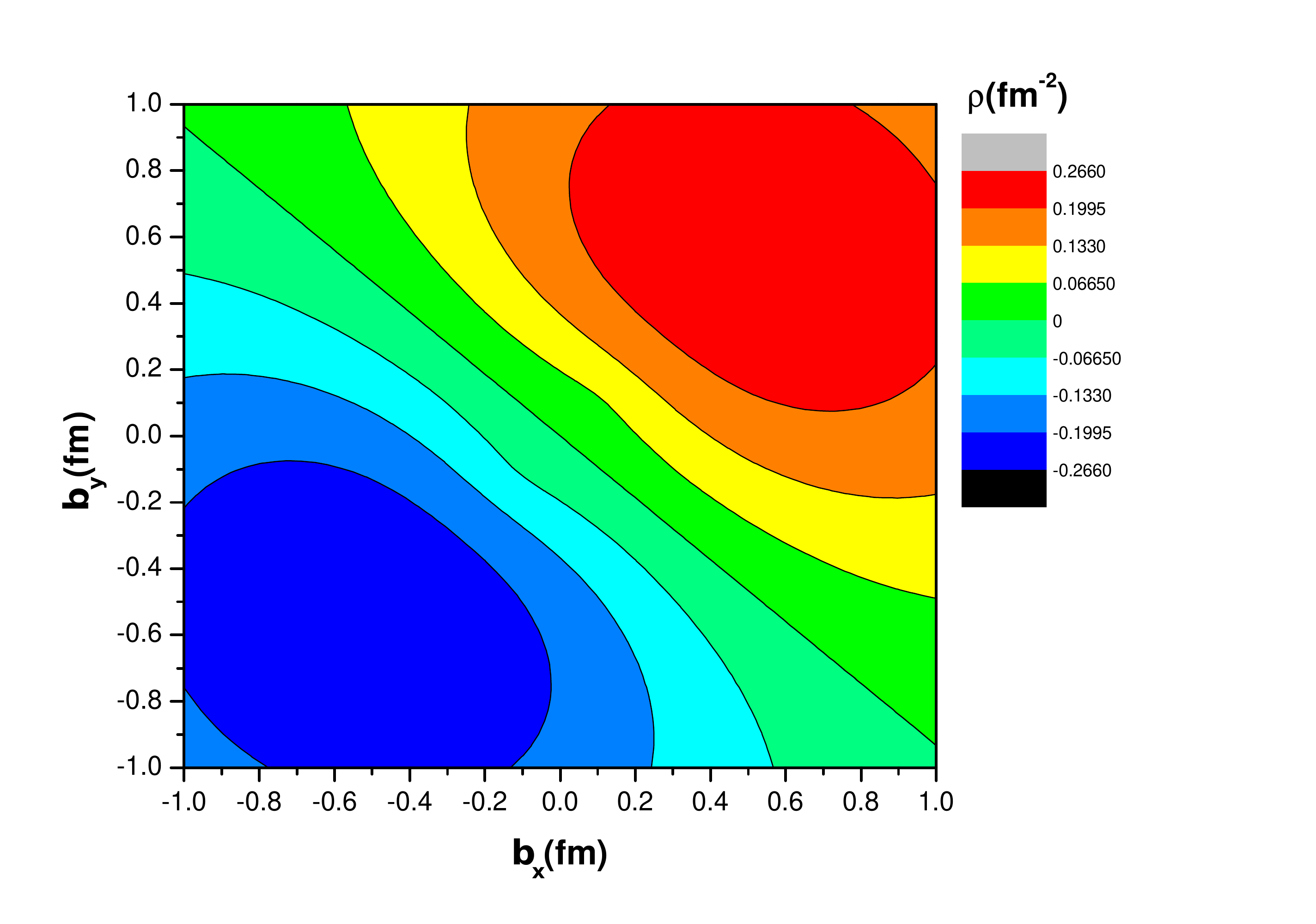}
 \endminipage\hfill
  \minipage{0.42\textwidth}
  \includegraphics[width=10cm]{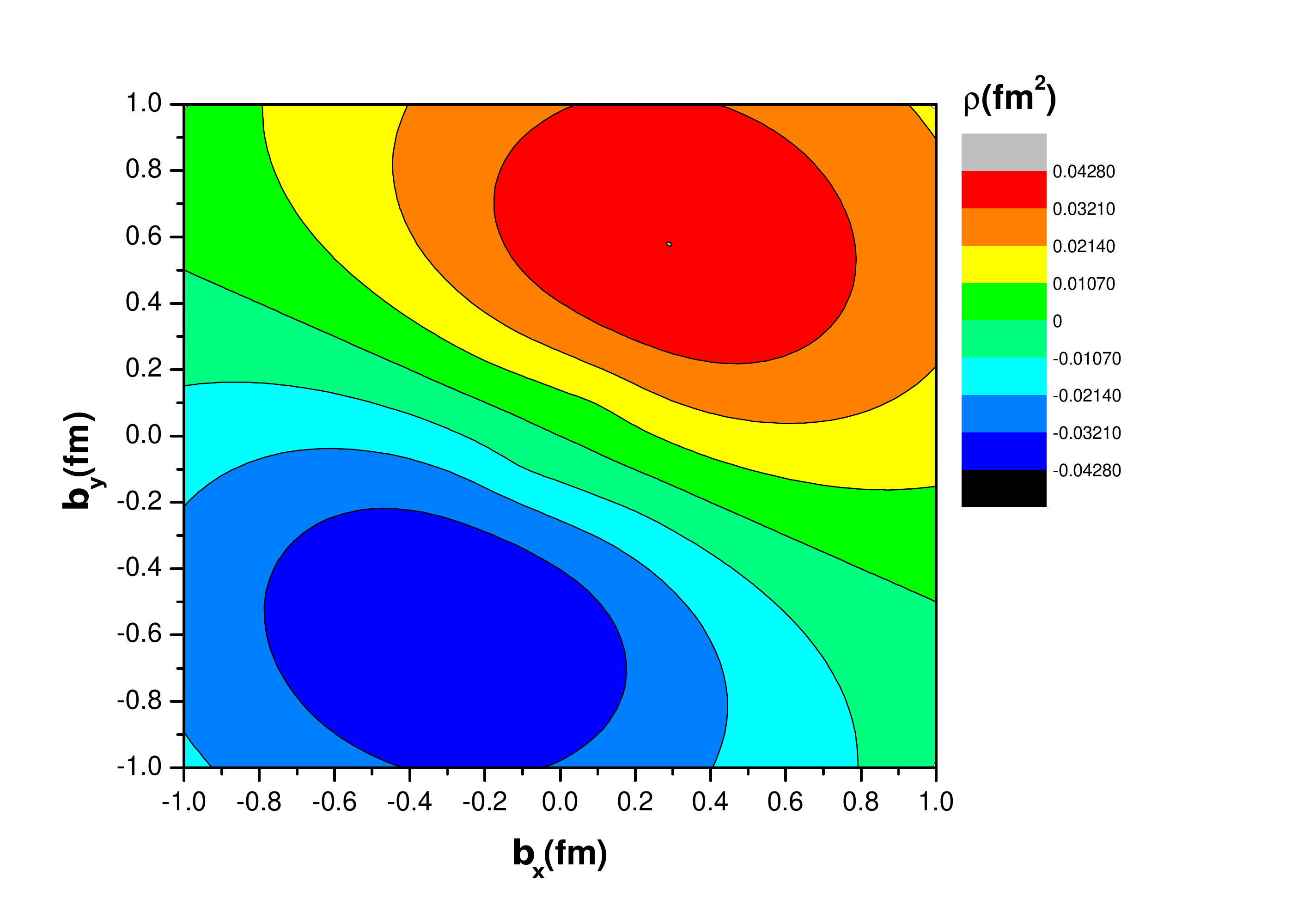}
  \endminipage\hfill
   \caption{The total dipole  $\frac{1}{2}[S_j b_i \mathcal{E'}- s_i b_j (\mathcal{E'}_T+2 \tilde{\mathcal{H}}'_T)/M]$ for $\hat{x}$ polarized quarks in a proton polarized in $\hat{y}$ direction for the up (left panel) and down (right panel) quarks.}
\label{total_dipole_chiral_even_and_odd}
\end{figure}

\begin{figure}[!]
\minipage{0.42\textwidth}
   \includegraphics[width=10cm]{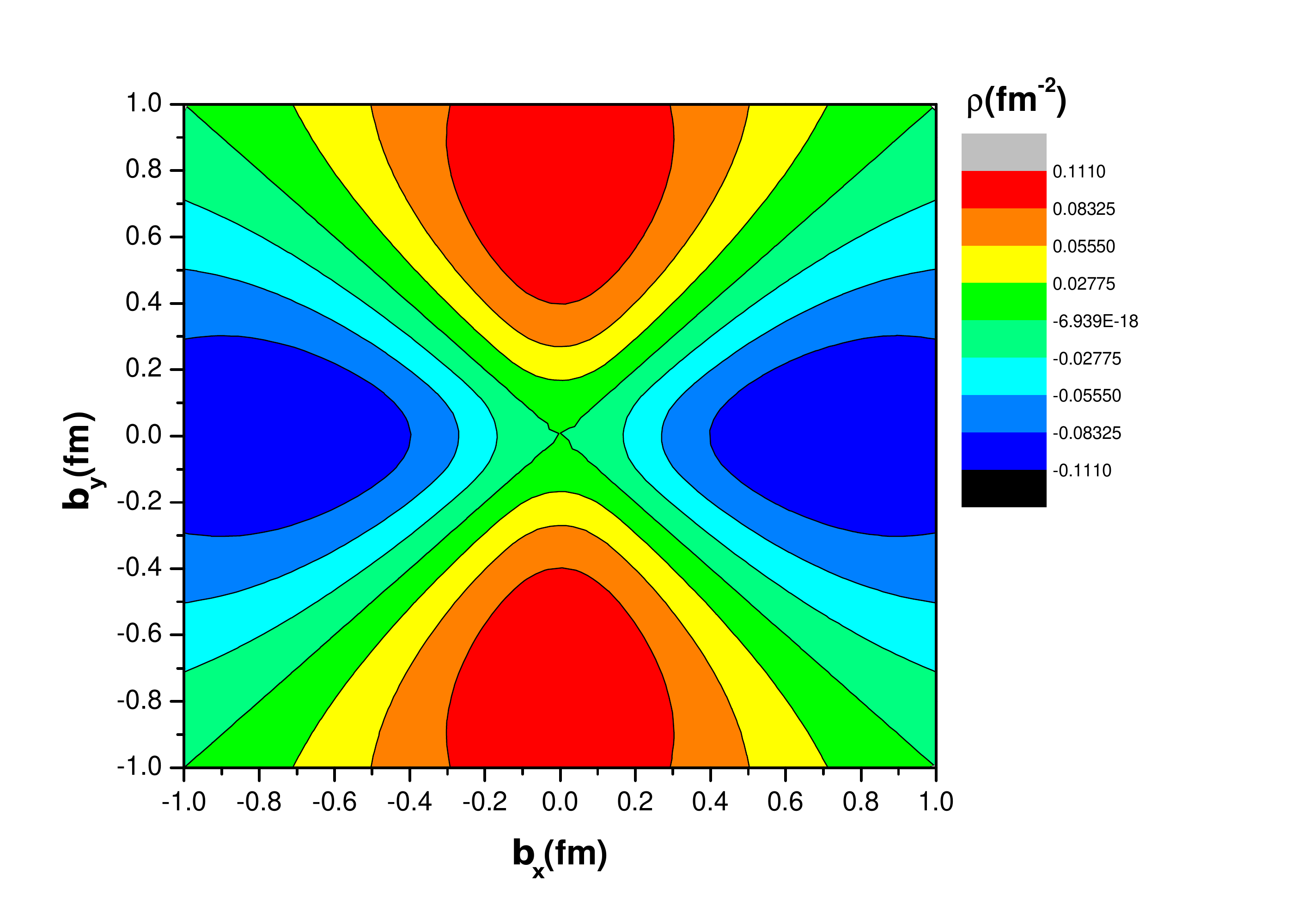}
 \endminipage\hfill
  \minipage{0.42\textwidth}
  \includegraphics[width=10cm]{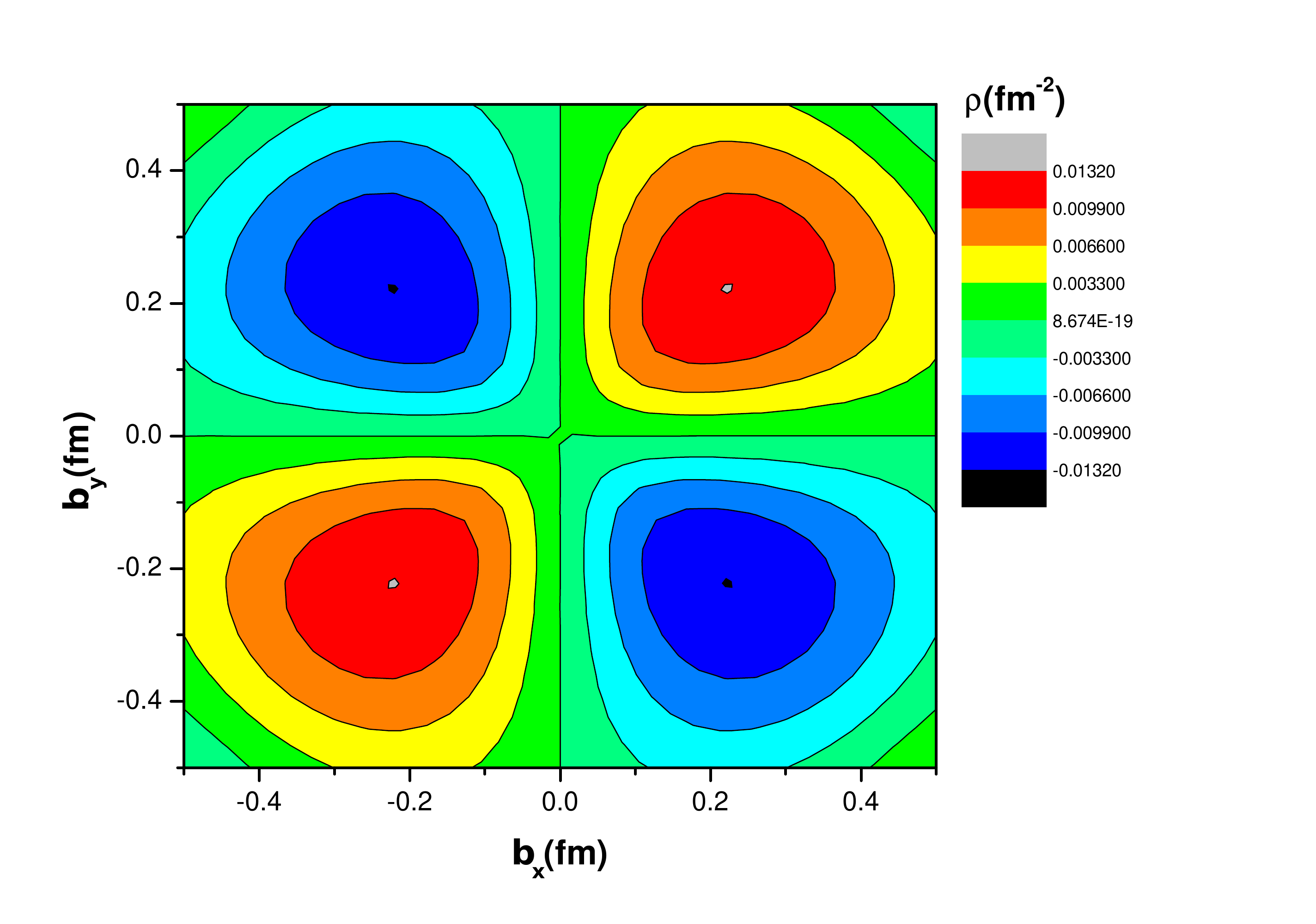}
  \endminipage\hfill
   \caption{The quadrupole contribution $s_i S_j b_i b_j \tilde{\mathcal{H}}''_T/M^2 $ for $\hat{x}$ polarized quarks in a proton polarized in $\hat{y}$ direction for the up (left panel) and down (right panel) quarks.}
\label{quadrupole}
\end{figure}

In Fig. \ref{dipole_chiral_even_x_axis}, \ref{total_dipole_chiral_even_and_odd} and \ref{quadrupole} we
have presented the results for dipole $\frac{1}{2} S_j b_i \mathcal{E'}/M$, total dipole $\frac{1}{2}[S_j b_i \mathcal{E'}-
s_i b_j (\mathcal{E'}_T+2 \tilde{\mathcal{H}}'_T)/M]$  and the quadrupole  $s_i S_j b_i b_j
\tilde{\mathcal{H}}''_T/M^2 $
contributions for $\hat{x}$ polarized quarks when the proton is transversely
polarized in the $\hat{y}$  direction. The distortion due to the dipole contribution
$\frac{1}{2} S_y b_x \mathcal{E'}$ in Fig. \ref{dipole_chiral_even_x_axis} gets rotated
with respect to the results shown in Fig. \ref{dipole_chiral_even_E} and the
total dipole contribution in Fig. \ref{total_dipole_chiral_even_and_odd} is
obtained from the second dipole term considered in Fig.
\ref{et+2ht_spin_density} with the additional factor of $S_j b_i \mathcal{E'}$. The result is sizeable and it is observed that the
density is larger for up quark than for the down quark. However, for the quadrupole term $s_i S_j b_i b_j
\tilde{\mathcal{H}}''_T/M^2 $, the result for the up quark is well spread over the plane whereas for the down quark the distribution is spread in almost half of the region as compared to the  up quark. Extensive work has been done in the light-front constituent
quark model \cite{pasquini} where the first moment of spin densities have been studied for
valence quarks. However, in the present work we have studied the spin densities
over a fixed value of $x$. One can further improve the results by considering the
meson cloud of the nucleon at the hadronic scale by including its contribution
in the evolution \cite{pasquinimeson,pasquinimeson1}.
\section{Summary and Conclusions \label{summ}}
In the present work, we have studied the chiral odd GPDs in the impact parameter space. We have considered a model with the
quark-proton scattering amplitudes at leading order with proton-quark-diquark vertices. It corresponds to a two body process consisting of a struck quark and
a diquark state. In order to obtain the explicit contributions from up and down quarks, we have considered  both the scalar (spin-0) and the axial-vector (spin-1) configurations for the diquark. Using the quark-proton helicity amplitudes, we have calculated the chiral odd GPDs for the case of zero skewness when the momentum transfer is in the totally transverse direction. In addition to this, we have also studied the spin densities for the up and down quarks for monopole, dipole and quadrupole contributions for unpolarized and polarized quarks in unpolarized and polarized proton.

For the case when unpolarized quarks are present in the unpolarized proton
the density distributions for the monopole $\mathcal{H}$ and dipole $\frac{-1}{2}s_i b_j (\mathcal{E'}_T+2 \mathcal{\tilde{H}}_T)$ terms are found to be larger for the up and down quarks and when we take the contributions from both the terms, the density distribution gets distorted in the plane.
Similarly, when we consider the monopole contributions for an unpolarized proton and the dipole
contribution from a transversely polarized proton, the density distribution
again gets distorted. We have also obtained the results for the polarized
quarks in the polarized proton for the monopole $\frac{1}{2} s_i S_i
(\mathcal{H}_T- \Delta_b \tilde{\mathcal{H}}'_T/{4 M^2})$ and the quadrupole $\frac{1}{2} s_i S_i (b_i^2 - b_j^2) \tilde{\mathcal{H}}''_T/ M^2$ contributions. Here again we find that the sign flip for the up and down quarks in the monopole and quadrupole contributions which is due to the different sign obtained for the $\mathcal{H}_T$ and $\tilde{\mathcal{H}}_T$ in the model. We have also considered
the $\hat{x}$-polarized quarks in the $\hat{y}$-polarized proton and the spin distribution is rotated with respect to the results obtained for
unpolarized quarks in unpolarized proton. The shift obtained here is however in the
same direction which leads to the same sign of the
magnetic moment of the up and down quarks. Similar results are obtained for the
case of quadrupole contributions. The spin densities provide a complete description
of the spin structure of the nucleon and its relation with TMDs could
be tested in future experiments.

\end{document}